\documentclass[twocolumn,preprintnumbers,amsmath,amssymb,superscriptaddress,nofootinbib,longbibliography]{revtex4-1}

\usepackage{graphicx}
\usepackage{bm}
\usepackage{dsfont}
\usepackage[usenames,dvipsnames]{xcolor}
\usepackage{pstricks}
\usepackage[tight]{subfigure}
\usepackage{verbatim}
\usepackage{units}
\usepackage{multirow}
\usepackage{enumitem}
\usepackage{mathrsfs}
\usepackage{leftidx}
\usepackage{xspace}
\usepackage{mathtools}
\usepackage{amsfonts,amssymb,amsmath}

\usepackage[a4paper]{hyperref}
\hypersetup{colorlinks=true,linktoc=all,linkcolor=blue,breaklinks=true,citecolor=blue,urlcolor=blue}

\usepackage[english]{babel}
\newcommand{\via}{\emph{via}\xspace}
\newcommand{\eg}{\emph{e.g.}\xspace}
\newcommand{\ie}{\emph{i.e.}\xspace}
\newcommand{\cf}{\emph{cf.}\xspace}
\newcommand{\etal}{\emph{et al.}\xspace}

\newcommand{\sA}{\text{A}}
\newcommand{\sB}{\text{B}}
\newcommand{\kB}{k_\text{B}}
\newcommand{\Vb}{V_\text{b}}
\newcommand{\acVb}{\delta V_\text{b}}
\newcommand{\dcVb}{\overline{V}_\text{\!b}}
\newcommand{\perT}{\mathcal{T}}

\newcommand{\opi}{\skew{2}{\hat}{i}}
\newcommand{\opI}{\skew{3.5}{\hat}{I}}
\newcommand{\auxF}{\mathcal{F}}
\newcommand{\FDeff}{\widetilde{f}}

\newcommand{\RR}{\text{R}}
\newcommand{\LL}{\text{L}}
\newcommand{\noise}{\mathcal{P}}
\newcommand{\en}{\mathcal{E}}
\renewcommand{\cap}{\text{cap}}

\newcommand{\avI}{\bar{I}}

\newcommand{\acU}{\delta U}
\newcommand{\dcU}{\overline{U}}

\newcommand{\acVg}{\delta V_\text{g}}

\newcommand{\Lor}{\mathscr{L}}

\newcommand{\slev}{\sigma_\text{lev}}
\newcommand{\scap}{\sigma_\text{cap}}
\newcommand{\scape}{\sigma_\text{cap}^{(\text{e})}}
\newcommand{\scaph}{\sigma_\text{cap}^{(\text{h})}}
\newcommand{\sloc}{\sigma_\text{loc}}

\newcommand{\loc}{\text{loc}}
\newcommand{\gate}{\text{g}}

\newcommand{\taug}{\tau_\text{g}}
\newcommand{\tauRC}{\tau_{RC}}

\newcommand{\Ctot}{C_\mu}
\newcommand{\Cq}{C_\text{q}}

\newcommand{\sC}{\text{C}}

\newcommand{\tleve}{t_\text{lev}}
\newcommand{\tcap}{t_\text{cap}}
\newcommand{\tcape}{t_\text{cap}^{(\text{e})}}
\newcommand{\tcaph}{t_\text{cap}^{(\text{h})}}

\newcommand{\tloce}{t_\text{loc}^{(\text{e})}}
\newcommand{\tloch}{t_\text{loc}^{(\text{h})}}
\newcommand{\tel}{t^{(\text{e})}}
\newcommand{\tho}{t^{(\text{h})}}

\newcommand{\capD}{\widetilde{D}}

\newcommand{\Eres}{E_\text{res}}

\newcommand{\Eresel}{E_\text{res}^{(\text{e})}}
\newcommand{\Eresk}{E_\text{res}^{(k)}}
\newcommand{\Eresj}{E_\text{res}^{(j)}}

\newcommand{\elec}{\text{e}}
\newcommand{\hole}{\text{h}}

\renewcommand{\tt}{\tilde{t}}
\newcommand{\tE}{\widetilde{E}}

\newcommand{\funA}{\mathcal{A}_\Gamma}
\newcommand{\funB}{\mathcal{B}_\Gamma}

\newcommand{\hgf}{\prescript{}{2}{F}_1^{}}
\newcommand{\tildehgf}{\prescript{}{2}{\widetilde{F}}_1^{}}
 
 \newcommand{\alpG}{\alpha_\Gamma}
 
 \newcommand{\alpen}{\alpha_\en}
 
 \newcommand{\sym}{\text{s}}
 \newcommand{\asym}{\text{as}}
 
 \newcommand{\lev}{\text{lev}}
 
 \newcommand{\auxI}{\mathcal{J}}
 
\usepackage{hyperref}

\AtBeginDocument{%
    \newwrite\bibnotes
    \def\bibnotesext{Notes.bib}
    \immediate\openout\bibnotes=\jobname\bibnotesext
    \immediate\write\bibnotes{@CONTROL{REVTEX41Control}}
    \immediate\write\bibnotes{@CONTROL{%
    apsrev41Control,author="08",editor="1",pages="1",title="0",year="1"}}
     \if@filesw
     \immediate\write\@auxout{\string\citation{apsrev41Control}}%
    \fi
}%

\begin{document}
	
	
	\title{Minimal-excitation single-particle emitters:\\ A comparison of charge and energy transport properties}
	
	\author{Nastaran Dashti}
	\affiliation{Department of Microtechnology and Nanoscience (MC2), Chalmers University of Technology, S-412 96 G\"oteborg, Sweden}

	\author{Maciej Misiorny}
	\affiliation{Department of Microtechnology and Nanoscience (MC2), Chalmers University of Technology, S-412 96 G\"oteborg, Sweden}

	\author{Sara Kheradsoud}
	\affiliation{Department of Physics, Lund University, S-221 00 Lund, Sweden}

	\author{Peter Samuelsson}
	\affiliation{Department of Physics, Lund University, S-221 00 Lund, Sweden}
		
	\author{Janine Splettstoesser}
	\affiliation{Department of Microtechnology and Nanoscience (MC2), Chalmers University of Technology, S-412 96 G\"oteborg, Sweden}
	
	\date{\today}
	
	\begin{abstract}
	We investigate different types of time-dependently driven single-particle sources whose common feature is that they produce pulses of integer charge and minimally excite the Fermi sea. 
	These sources are: a slowly driven mesoscopic capacitor, a Lorentzian-shaped time-dependent bias voltage, and a local gate-voltage modulation of a quantum Hall edge state. They differ by their specific driving protocols, \eg, they have a pure ac driving or a driving with a dc component.
	In addition, only in the first of these setups, strong confinement leading to a discrete energy spectrum of the conductor, is exploited for the single-particle emission. 
	Here, we study if and how these basic differences impact transport properties. Specifically, we address time- and energy-resolved charge and energy currents, as well as their zero-frequency correlators (charge-, energy- and mixed noise), as they are frequently used to characterize experiments in quantum optics with electrons. 
	Beyond disparities due to a different number and polarity of particles emitted per period, we in particular identify differences in the impact, which temperature has on the observables for sources with and without energy-dependent scattering properties. We  trace  back  these characteristics to a small set of relevant parameter ratios.
	\end{abstract}

	\maketitle

\section{Introduction}

In recent years,  transport properties, such as charge currents and their correlations have been employed as analysis tools in quantum optics with electrons~\cite{Bocquillon2014Jan,Bauerle2018Apr}. 
In this field, single electrons are injected into a conductor by time-dependently operated single-particle sources~\cite{Blumenthal2007Apr,Feve07,McNeil2011Sep,Hermelin2011Sep,Dubois13,Fletcher2013Nov,Ubbelohde2014Dec,vanZanten2016Apr} and sent on quantum point contacts (QPCs) acting as beam splitters, possibly even employing quantum Hall edge states as electronic waveguides. This combination of tuneable device elements allows for implementation of optics-like experiments for electrons and has triggered a great number of further proposals, like the Mach-Zehnder~\cite{Haack11,Juergens2011Oct,Hofer2014Dec,Rossello2015Mar,Dasenbrook2015Oct}, the Hanbury Brown-Twiss~\cite{Grenier11,Bocquillon2012May,Bocquillon2013May,Jullien14,Thibierge2016Feb} or the Hong-Ou-Mandel interferometer~\cite{Ol'khovskaya2008Oct,Moskalets2011Jan,Jonckheere12,Bocquillon2013Mar,Wahl2014Jan,Freulon2015Apr,Marguerite2016Sep,Marguerite2017Oct}, with additional complexities arising for example from the presence of the Fermi sea or from Coulomb interactions between electrons~\cite{Bocquillon2013May,Wahl2014Jan,Iyoda2014May,Freulon2015Apr,Marguerite2016Sep,Cabart2018Oct}. The scope of the proposed or already conducted experiments ranges from the signal analysis of the single-particle source~\cite{Grenier11,Gabelli13,Jullien14,Thibierge2016Feb,Marguerite2017Oct,Kashcheyevs2017} to solid-state based entanglement protocols~\cite{Splettstoesser2009Aug,Sherkunov2012,Vyshnevyy2013Apr,Hofer2013Dec,Inhofer2013Dec,Dasenbrook2015Oct}.
In addition to the study of charge currents, also energy currents~\cite{Moskalets2009Aug,Rossello2015Mar} and their correlations have recently come into  focus~\cite{Battista13,Battista14,Moskalets2014May,Battista14a,Moskalets2017Mar,Vannucci2017Jun,Dashti2018Aug,Ronetti2018Nov}, since they possibly yield further information about the spectral properties of the setup of interest.

\begin{figure}[b!!!]
\includegraphics[scale=1]{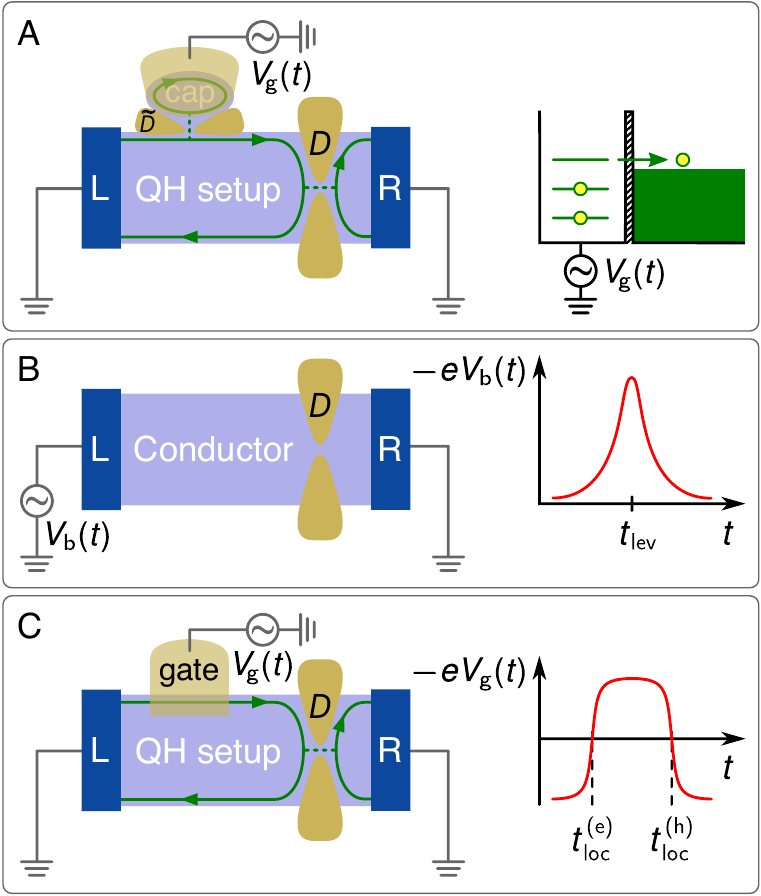}
	\caption{\label{fig_setup}
	Emission schemes of single particles:
	(A) a driven mesoscopic capacitor (cap) in a quantum-Hall (QH) setting, with arrowed lines indicating a single conducting edge channel; 
	(B) a voltage-biased conductor;
	and (C) a QH edge state, locally modulated by a gate potential.
	Currents and their correlations are detected at contact~R.
}
\end{figure}

The single-particle sources, which are used to emit electrons (and holes) into an electronic optics setup can be of very different nature~\cite{Blumenthal2007Apr,Feve07,McNeil2011Sep,Hermelin2011Sep,Dubois13,Fletcher2013Nov,Ubbelohde2014Dec,vanZanten2016Apr}. Here, we focus on those sources, which can be designed to emit single particles as minimal, noiseless excitations of the Fermi sea. 
More than ten years ago, a pure ac single-particle source based on a driven mesoscopic capacitor~\cite{Pretre96,Buttiker93a} in the quantum Hall regime was first implemented~\cite{Feve07}, and it has since then been used in a variety of experimental setups (see setup A in Fig.~\ref{fig_setup}). 
On the other hand, a very different design of a single-electron source has been realized~\cite{Dubois13,Gabelli13} by shaping the time-dependent bias voltage across a conductor in a specifically chosen Lorentzian form~\cite{Levitov96,Keeling06,Keeling08,Sherkunov2009Jul,Dubois13Theo} (see setup B in Fig.~\ref{fig_setup}). 
Finally, a third setup was recently suggested~\cite{Misiorny2018Feb}, in which the local gate modulation  of an edge state allows for noiseless single-particle emission on top of the Fermi sea (see setup C in Fig.~\ref{fig_setup}). 
While all these three types of sources enable a noiseless emission of single particles, their operation principles and specific designs differ largely.\footnote{Other single-particle sources not addressed here inject particles far above the Fermi sea.}
In experiments, which can possibly involve several synchronized sources among other device elements, it is hence of vital importance  to understand, how the basic source properties reflect in the various relevant observables and when results are expected to be source-independent.

In this paper, we analyze how the properties of the different sources, schematically depicted in Fig.~\ref{fig_setup}, impact the time-resolved charge and energy currents as well as the spectral current.
Furthermore, inclusion of a QPC into the setup allows us to study the zero-frequency charge-current and energy-current noise of the injected signals, as well as the mixed noise between charge and energy currents. 
For this purpose, we employ a scattering matrix approach, which is valid as long as the effective Coulomb interaction between electrons is weak. This is a reasonable assumption, since the Coulomb interaction in the only confined studied element (the mesoscopic capacitor) is typically screened by top gates.
Interaction effects arising during the propagation along edge states~\cite{Bocquillon2013May,Wahl2014Jan,Iyoda2014May,Freulon2015Apr,Marguerite2016Sep,Cabart2018Oct} are not addressed here; we instead focus on the properties of the sources alone in order to disentangle them from other occurring effects.

We find that the zero-temperature limit of currents and noises are to a large extent similar for the three emission schemes under consideration, when operated in a regime of well-separated single-particle pulses, except for the number and type of emitted particles (electrons and holes). 
The situation changes at finite temperatures. In this case, it depends strongly on the role which the energy dependence of the scattering matrix describing the setup has on the observable under consideration.

The paper is organized as follows: 
In Sec.~\ref{sec_model}, we introduce the three different emission protocols and define relevant parameter regimes for their operation. 
Next, the time-resolved charge and energy currents are analyzed in Sec.~\ref{sec_time_resolved}, followed by a study of the spectral current in Sec.~\ref{sec_spec}. 
Finally, the zero-frequency noise of charge and energy currents, as well as their mixed noise, are discussed in Sec.~\ref{sec_noise}. 
Extensive Appendices~\ref{app_general_exp}-\ref{app_deltat_contrib} contain general definitions, full analytical results and crucial elements of their rigorous derivations.

\section{Single-particle emission schemes}\label{sec_model}

In this section, we introduce the three different single-particle emission schemes  that we are going to compare: 
\begin{itemize}[leftmargin=*]
\item
a gate-voltage driven mesoscopic capacitor embedded in a quantum-Hall setup (setup A),
\item
a coherent conductor with a time-dependent Lorentz\-ian-shaped voltage-bias driving (setup B),
\item
a quantum Hall (QH) edge state locally modulated by a gate voltage with a smooth-box shape (setup C).
\end{itemize}
A schematic illustration of these three schemes is presented in Fig.~\ref{fig_setup}.
In the operation regimes, which we consider here, the common property of these three sources is that they emit a noiseless stream of single particles, as a minimal excitation just on top of the Fermi sea. In all three cases, single-particle excitations are sent from the left side onto a QPC with a single open channel with transmission~$D$, which we assume to be energy-independent. The conductors are in contact with two reservoirs~L~(left) and~R (right). The occupation of electronic states in the two reservoirs is characterized by Fermi functions 
\mbox{$
	f_\alpha(E)
	=
	\big\{
	1+\exp[(E-\mu_\alpha)/(\kB T)]
	\big\}^{-1}
$},
with electrochemical potentials $\mu_\alpha$,\footnote{In the case of time-dependent driving of the electrochemical potential (setup B), only the constant part of $\mu_\alpha$ enters the Fermi function, see also the appendix of Ref.~\cite{Battista14}.} temperature $T$, and the Boltzmann constant $k_\text{B}$. We fix the electrochemical potential of the right reservoir as the energy reference~\mbox{$\mu_\RR=0$}. This is useful since we assume that charge and energy currents, as well as their correlations, are always detected in contact  R. In order to actually evaluate charge and energy currents, as well as their noise, we use a Floquet scattering matrix approach, see, \eg,  Ref.~\cite{Moskalets_book}, as well as Appendix~\ref{app_general_exp} and \ref{app_general_results} for definitions of all observables in terms of scattering matrices and general results. The description of the three time-dependently driven setups in terms of  scattering matrices or excitation amplitudes is discussed in the following.

\subsection{Time-dependently driven mesoscopic capacitor}\label{sec_model_capa}

The conductor of setup~A is in the quantum Hall regime, so that transport takes place \via chiral edge states---marked in Fig.~\ref{fig_setup} as dark green lines with arrows indicating the propagation direction. Setup~A contains a mesoscopic capacitor  (designated with `cap') operating as a single-particle source, which consists of a small region, weakly coupled to the edge by an additional QPC with transmission probability~$\capD$.  Such a setup was first realized by F\`{e}ve~\etal~\cite{Feve07}. There is no voltage bias applied, \mbox{$\mu_\LL=\mu_\RR\equiv0$}.
The capacitor region is driven at frequency~$\Omega$ by a time-dependent potential~\mbox{$U(t)=\dcU-\acU\sin(\Omega t)$} induced by a top gate, so that charge is emitted into the chiral edge state and impinges onto the central QPC. %
The time-dependently driven mesoscopic capacitor is described by a Floquet scattering matrix~$S(E_n,E)$, which depends on the energy $E$ of incoming states as well as on the transferred energy $n\hbar\Omega=E_n-E$, see, \eg, Ref.~\cite{Moskalets_book}.
Due to the finite size of the confined capacitor region, the scattering matrix becomes dependent on energy. In the case we consider here, \mbox{$\widetilde{D}\ll1$}, the capacitor spectrum is quasi-discrete, with a strongly energy-dependent scattering matrix.

We are interested in a regime, where the mesoscopic capacitor emits well-separated pulses of single-particle (electron- and hole-) excitations close to the Fermi energy. This happens under the following three conditions: 
First of all, (i) the driving needs to be sufficiently slow, which essentially means that the adiabatic-response condition, 
\begin{equation}\label{eq_adiab_cond}
	\frac{\Omega\tau}{\capD}
	\ll
	1
	,
\end{equation} 
needs to be  fulfilled. Here, $\tau$ stands for the time an electron needs to make one turn around the capacitor cavity.
Furthermore, (ii) the condition that current pulses are well-separated in time is expressed as 
\begin{equation}\label{eq_res_cond}
	\Omega\scap\ll1
	,
\end{equation}
with the temporal width of the pulses (at zero-tem\-pe\-ra\-ture) given by $\sigma_\cap$. Thus, this width needs to be much smaller than the period of the driving \mbox{$\perT=2\pi/\Omega$}. 
For the same reason, (iii) it needs to be guaranteed that
\begin{equation}\label{eq_temp_cond}
	\kB T
	\ll
	|e\acU|
	,
\end{equation} 
in other words, that the excess energy provided by a finite temperature of the contacts does not exceed the driving amplitude.
However, pulses could still overlap under these last two conditions, if the constant part of the gate potential is chosen in an unfavorable way, thereby hindering the energy level of the capacitor from fully crossing the Fermi energy of the contacts during the driving cycle. For simplicity, we hence set \mbox{$\dcU\equiv0$}.
Note that for very large driving amplitudes, that is, larger than the level spacing of the capacitor \mbox{$|e\acU|>\Delta$}, more than one electron and more than one hole can be emitted, see, \eg,  Ref.~\cite{Moskalets2008Feb}.  Here, we avoid this situation by choosing \mbox{$|e\acU|<\Delta$} and concentrate on the case of single-electron emission\footnote{%
Note that $\Delta$ does not directly enter our model: the condition \mbox{$|e\acU|<\Delta$} allows us to consider a single-level spectrum for the capacitor.}. 

Under these conditions, the energy-dependent scattering matrix takes the form
\begin{equation}\label{eq_S0_resonant_enrep}
\hspace*{-4pt}
	S_\cap(E_{n},E)
	\!=\!
	\left\{\begin{array}{ll}
	\!\!
	-
	2\Omega\scap
	\,
	e^{-n\Omega\scap}
	\,
	e^{in\Omega\tcape(E)}
	&\! n>0
	,
\\[3pt]
	\!\!1 &\! n=0
	,
\\
	\!\!
	-
	2\Omega\scap
	\,
	e^{n\Omega\scap}
	\,
	e^{in\Omega\tcaph(E)} &\! n<0
	.
	\end{array}\right. 
	\!\!
\end{equation}
Here, the width of the pulses, $\sigma^{(\elec/\hole)}_\text{cap}$, and the energy-dependent emission times, $t^{(\elec/\hole)}_\text{cap}(E)$, for electrons~(e) and holes~(h), are related to the characteristics of the capacitor design and the driving potential as
\begin{gather}\label{eq_tcap_energy}
	\tcape(E)
	 =
	\frac{E}{e\acU\Omega}  =\frac{(2m+1)\pi}{\Omega}-\tcaph(E)
	,
\\\label{eq_scap_energy}
	\scap  \equiv  \scaph	 
	= 
	\frac{\capD\hbar}{2e\acU\tau\Omega} =-\scape
	\ ,
\end{gather}
with \mbox{$m\in \mathbb{Z}$}. Here, $m$ is only required to account for emission in different driving periods; in what follows, we focus on one emission period only and fix the order in which electrons and holes are emitted, such that we can take $m=0$. 
Equations~\eqref{eq_tcap_energy} and \eqref{eq_scap_energy} agree with the limit for small $E/|e\delta U|$---relevant if the condition~(iii) as given in Eq.~\eqref{eq_temp_cond} is fulfilled---of the full expressions for emission times and pulse widths obtained in Ref.~\cite{Splettstoesser2008Nov}.

For the analysis below it is helpful to write the scattering matrix in a mixed time-energy representation, $S(t,E)$, related to the previously introduced Floquet scattering matrix $S(E_n,E)$ \via 
\begin{equation}\label{eq_Smatrix_mixed}
	S_\cap(E_n,E)
	=
	\int_0^\mathcal{T}\!\frac{dt}{\mathcal{T}}
	\,
	S(t,E)
	\,
	e^{in\Omega t}.
\end{equation}
Then, the resonant emission of electrons and holes in the vicinity of the emission times $\tcap^{(j)}$ becomes particularly clear
\begin{equation}\label{eq_t_resonant}
	S_\cap(t,E)
	=
	\sum_{j=\elec,\hole}
	\frac{t-\tcap^{(j)}(E)-i\scap^{(j)}}{t-\tcap^{(j)}(E)+i\scap^{(j)}}
	.
\end{equation}
Equivalently, one can rewrite the scattering matrix above in terms of relevant emission energies. This formulation emphasizes that an energy level of the discrete spectrum of the capacitor with width $\Gamma$ has to be energetically accessible in order to allow for the emission of electrons and holes,
\begin{equation}\label{eq_E_resonant}
	S_\cap(t,E)
	=
	\sum_{j=\elec,\hole}
	\frac{E-\Eres^{(j)}(t)-i\Gamma}{E-\Eres^{(j)}(t)+i\Gamma}
	,
\end{equation}
with the level broadening~$\Gamma$ and the emission energies $\Eres^{(j)}(t)$ defined as
\begin{gather}\label{eq_tE}
	\Eres^{(\elec)}(t)
	=
	e\delta U\Omega t
	=
	e\delta U\pi
	-
	\Eres^{(\hole)}(t)
	,
\\\label{eq_gamma}
	\Gamma
	=
	\frac{\hbar\widetilde{D}}{2\tau}
	= 
	e\delta U\Omega\sigma_\text{cap}	
	.
\end{gather}
The adiabatic condition (i) and the resonance condition~(ii), given in Eqs.~(\ref{eq_adiab_cond}) and (\ref{eq_res_cond}), respectively, can then be alternatively formulated in terms of energy scales only,
\begin{equation}\label{eq_cond_2}
	\hbar\Omega\ll2\Gamma 
	\quad
	\text{and} 
	\quad
	\hbar\Omega\ll2\mathcal{E}
	.
\end{equation}
Here, we have introduced the energy, 
\begin{equation}\label{eq_en_def}
	\mathcal{E}
	=
	\hbar/(2\sigma_\cap)
	,
\end{equation}
which will later turn out to characterize the energy carried by an emitted pulse. Note that while the energy carried by the pulse $\mathcal{E}$ is a relevant concept for the single-particle emission from all three setups, the level width $\Gamma$ is a parameter that only describes the mesoscopic capacitor with its discrete spectrum arising due to the confinement. In particular, $\Gamma$ is independent of the driving.

\subsection{Lorentzian bias voltage}\label{sec_model_leviton}

The conductor of setup~B does not require a magnetic field. In order to produce controlled single-particle excitations in setup~B, a time-dependent potential is applied to the left contact
\mbox{$
	\mu_\text{L}(t)/(-e)
	\equiv
	\Vb(t)
	=
	\acVb(t)+\dcVb
$}.
Here, $\acVb(t)$ and $\dcVb$ stand for the pure ac and dc components of the driving potential~$\Vb(t)$, respectively, and $-e$ (with \mbox{$e>0$}) is the electron charge.
The application of a time-dependent voltage~$\Vb(t)$ leads to a spread in energy of the electronic states injected from the left contact and impinging on the scatterer~\cite{Pretre96}. Therefore, one has to relate the creation and annihilation operators $\hat{a}^\dagger_\LL(E)$ and $\hat{a}_\LL(E)$, for current-carrying states entering the conductor after having been subject to the time-dependent driving, to those deep in the contact, referred to as $\hat{a}^{0\dagger}_\LL(E)$ and $\hat{a}^0_\LL(E)$. Importantly, only the latter operators obey Fermi-Dirac statistics, 
\mbox{$
	\big\langle
	\hat{a}_\LL^{0\dagger}(E)
	\hat{a}_\LL^{0}(E^\prime)
	\big\rangle
	=
	\delta(E-E^\prime)
	f_\LL(E)
$.}
The relation between the operators $\hat{a}_\LL(E)$ and $\hat{a}_\LL^0(E)$ is 
\begin{equation}\label{eq_a_vs_a0}
	\hat{a}_\LL(E)
	=
	\sum_{n=-\infty}^{\infty}
	\!
	c_{\LL,n}
	\,
	\hat{a}_\LL^0(E_{-n})
	,
\end{equation}
with the amplitude~$c_{\LL,n}$, 
\begin{equation}\label{eq_cLm}
	c_{\LL,n}
	=
	\int_0^\mathcal{T}\!\frac{dt}{\mathcal{T}}
	\,
	e^{in\Omega t}
	\,
	e^{-i\varphi_\LL(t)}
	,
\end{equation}
basically  defined as the Fourier coefficient of the phase factor $\exp[-i\varphi_\LL(t)]$, and the time-dependent phase being
\begin{equation}
	\varphi_\LL(t)
	=
	-\frac{e}{\hbar}
	\int_0^t\!dt^\prime
	\,
	\acVb(t^\prime)
	.
\end{equation}
The amplitudes~$c_{\LL,n}$ determine the probability $|c_{\LL,n}|^2$ that $n$ Floquet energy quanta~$\hbar\Omega$ are emitted (\mbox{$n<0$}) or absorbed (\mbox{$n>0$}) in a scattering process. Note that in contrast to the scattering matrix of setup~A, these amplitudes given by Eq.~(\ref{eq_cLm}) are energy-independent.

Here, we are interested in a specific shape of the driving signal, namely, a periodically repeated Lorentzian-shaped bias,
\begin{equation}\label{eq_Lor_bias}
	\Vb(t) 
	= 
	V\perT
	\sum_{j=-\infty}^\infty
	\Lor_{\slev}\big(t-\tleve-j\mathcal{T}\big)
	,
\end{equation}
with the Lorentzian function~\mbox{$\Lor_y(x)=y\left[\pi(x^2+y^2)\right]^{-1}$}.
The voltage pulse is centered around~$\tleve$, which we refer to as the particle emission time, and has a full width at half maximum given by~$2\slev$. 
It has been theoretically predicted~\cite{Keeling06,Levitov96,Keeling08}, and later also verified in experiments~\cite{Dubois13,Gabelli13}, that under the condition \mbox{$V=\hbar\Omega/(-e)$}, such a Lorentzian-shaped driving signal leads to a noiseless emission of independent single-electron excitations (one per period), known as \emph{levitons}. This is the regime, which we will address in this paper, where equivalent to the condition given in Eq.~(\ref{eq_res_cond}), we require \mbox{$\Omega\slev\ll1$} to guarantee that pulses are well separated in time. In this resonant regime, the amplitudes $c_{\LL,n}$ take the form
\begin{equation}\label{eq_cLn_resonant}
	c_{\LL,n}
	=
	\left\{\begin{array}{ll}
	-
	2\Omega\slev
	\,
	e^{-n\Omega\slev}
	\,
	e^{i(n+1)\Omega\tleve}
	& n>-1
	,
\\[3pt]
	-e^{-\Omega\slev}
	& n=-1
	,
\\[3pt]
	0 & n<-1
	.
	\end{array}\right. 
\end{equation}
%

\subsection{Local time-dependent edge-state modulation}\label{sec_model_loc}

Setup~C combines some of the properties of the previously introduced setups~A and B, see Fig.~\ref{fig_setup}. Similar as in the mesoscopic capacitor setup, the conductor is in the quantum Hall regime and the driving is locally applied \via a pure ac gate voltage. However, no confinement is exploited here, instead---similar as in setup~B---it is the specifically chosen shape of the driving gate voltage which leads to the noiseless emission of particles (here both electrons and holes).
The scattering matrix describing the effect of the induced electric field on electrons traversing the gated region is given by
\begin{align}\label{eq_Sloc}
	S_\loc(t,E)
	=\ &
	e^{iE\taug/\hbar}
	e^{-i\varphi_\gate(t)}
\nonumber\\
	=\ &
	e^{iE\taug/\hbar}
	\sum_{n=-\infty}^\infty
	\,
	c_{\gate,n}	
	\,
	e^{-in\Omega t}
	.
\end{align}
Here, $\taug$ denotes the traversal time of an electron passing through the gated region. The amplitudes~$c_{\gate,n}$ are defined analogously to Eq.~(\ref{eq_cLm}), with
\begin{equation}
	\varphi_\gate(t)
	=
	-\frac{e}{\hbar}
	\int_{t-\taug}^t\!dt^\prime
	\,
	\acVg^\text{eff}(t^\prime)
	.
\end{equation}
The relation between the internal potential~$\acVg^\text{eff}(t^\prime)$ in the interacting region and the potential $\acVg(t)$ applied to the gate can in general be fairly non-trivial~\cite{Misiorny2018Feb}. Here, for simplicity, we focus on the adiabatic-response regime, requiring  \mbox{$\tauRC\ll\perT$}. 
The $RC$-time,
\mbox{$
	\tauRC=R\Ctot
$},
is the product of the B\"uttiker resistance \mbox{$R=h/(2e^2)$} and the total (electrochemical) capacitance $\Ctot$, which for a metallic gate is given by 
\mbox{$
	\Ctot^{-1}
	=
	C^{-1}
	+
	\Cq^{-1}
$}, 
with the purely electrostatic capacitance~$C$ and the quantum capacitance \mbox{$\Cq=\taug e^2/h$}. In the adiabatic-response limit, one derives \mbox{$\acVg^\text{eff}(t)=(\Ctot/\Cq)\acVg(t)$} and
\begin{equation}
	\varphi_\gate(t)
		=
	2\pi
	\frac{\Ctot}{(-e)}
	\acVg(t)\ .
\end{equation}
In order to obtain the noiseless emission of electrons and holes in this driving regime, the required ac gate potential is a smooth-box potential
\begin{multline}
	\acVg(t)
	=
	\frac{\acVg}{2\pi}
	\text{Re}
	\bigg\{
	i
	\,
	\text{ln}
	\bigg[
	\frac{
	\sin\!\big(\Omega[t-\tloce+i\sloc]/2\big)
	}{
	\sin\!\big(\Omega[t-\tloch+i\sloc]/2\big)
	}
	\bigg]
	\!
	\bigg\}
\\
	-
	\frac{\acVg}{2}
	,
\end{multline}
fulfilling the requirement \mbox{$\Ctot\acVg/(-e)=1$}. Here, we assume that the steps of the box are evenly distributed, \ie, $\tloch-\tloce=\perT/2$, meaning that the emission time of electrons and holes is half a period apart. Moreover, these pulses are well separated in time if \mbox{$\Omega\sloc\ll1$}. Then, the coefficients $c_{\gate,n}$ in Eq.~\eqref{eq_Sloc} read
\begin{equation}\label{eq_cgn_resonant}
	c_{\gate,n}
	=
	\left\{\begin{array}{ll}
	-
	2\Omega\sloc
	\,
	e^{-n\Omega\sloc}
	\,
	e^{in\Omega\tloce}
	& n>0
	,
\\[3pt]
	1 & n=0
	,
\\
	-
	2\Omega\sloc
	\,
	e^{n\Omega\sloc}
	\,
	e^{in\Omega\tloch} & n<0
	.
	\end{array}\right. 
\end{equation}

In what follows, we are going to discuss the currents of the sources and the related noise in different regimes, within the approximations fixed above. In order to compare the properties of the currents from the three sources, we take them to be characterized by equal widths \mbox{$\sigma\equiv\sigma_\cap=\sigma_\lev=\sloc$} and emission times, \mbox{$\tel\equiv t_\lev=\tel_\cap(E=0)=\tloce=0$} (focussing on one emission period, only). In our comparison, we are particularly interested in the influence of temperature~\cite{Moskalets2016Jan,
Moskalets2017Jul,Moskalets2017Oct}. For this reason, we define two dimensionless parameters
\begin{gather}
	\alpen
	:=
	\frac{\kB T}{\mathcal{E}}
	,
	\\
	\alpG
	:=
	\frac{\kB T}{\Gamma}
	,
\end{gather}
which are not fixed by the described conditions of adiabaticity and emission of separate pulses. They define the ratio of temperature with respect to the energy carried by the emitted pulses~($\alpen$), as well as to the driving-independent level width of the capacitor's energy spectrum~($\alpG$).

\section{Time-resolved currents}\label{sec_time_resolved}

In order to understand the relevant energy scales entering the time-resolved charge and energy currents emitted from the sources, it is helpful to consider the general expression for currents in the right contact
\begin{align}\label{eq_It_general}
\hspace*{-4pt}
	\renewcommand{\arraystretch}{1.35}
	\begin{pmatrix}
	I(t)
	\\
	I^\en(t)
	\end{pmatrix}	
	=\ & 
	\frac{D}{h}\sum_{n,\ell=-\infty}^{\infty}\int dE
	\renewcommand{\arraystretch}{1.35}
	\begin{pmatrix}
	-e
	\\
	E+\frac{(n+\ell)\hbar\Omega}{2}
	\end{pmatrix}	
\nonumber\\
	&\!\times
	\left[f_\text{L}(E)-f_\text{R}(E_n)\right]
	\int_0^\mathcal{T}\frac{dt'}{\mathcal{T}}e^{in\Omega(t-t')}
\nonumber\\
	&\!\times
	\int_0^\mathcal{T}\frac{dt''}{\mathcal{T}}e^{-i\ell\Omega(t-t'')}
	S^*(t',E)
	S(t'',E)
	;
\end{align}
see also Appendix~\ref{app_general_results}. For the driven mesoscopic capacitor (setup~A) and the locally modulated edge state (setup~C), the subscript of the Fermi functions can be dropped. Moreover, in order to describe the leviton emission (setup B), scattering matrices have to be replaced by \mbox{$S(t,E)\rightarrow\exp[-i\varphi_\LL(t)]$}. 

First of all, it can be noticed that in Eq.~(\ref{eq_It_general}), there is the energy dependence given by the scattering matrix itself. This energy dependence is always absent for the leviton emission and also for the locally modulated edge state the energy-dependence of the scattering matrix does not contribute to Eq.~\eqref{eq_It_general}. Besides, for the slowly driven mesoscopic capacitor, the scattering matrix becomes energy-independent on the energy-scale~$\mathcal{E}$ set by the driving, \cf~Eqs.~(\ref{eq_scap_energy}) and~(\ref{eq_en_def}). However, there is an additional energy dependence entering \via the Fermi functions. Due to this energy dependence on the scale of the temperature, we can replace \mbox{$S(t,E)\rightarrow S(t,0)$} for the capacitor setup only as long as  temperature does not by far exceed the energy scale $\mathcal{E}$, namely as long as $\alpha_\mathcal{E}\lesssim1$.

As a result, in the limits considered here, the time-resolved currents going along with the single-particle emission from the three sources have basically equivalent features when \mbox{$\alpha_\mathcal{E}\lesssim1$}. The only differences then stem from the fact that in one driving period one electron \emph{and} one hole are emitted from the capacitor and the locally modulated edge state, while the bias-driving results in the emission of particles of one polarity, only. In what follows, we discuss features of time-resolved currents for the leviton emission  and the locally modulated edge state always together with the low- to intermediate-temperature regime (\mbox{$\alpha_\mathcal{E}\lesssim1$}) of the mesoscopic capacitor, since the behavior is basically equivalent. At higher temperatures, the signal emitted from the mesoscopic capacitor is modified---in contrast to the leviton and the signal emitted from the locally modulated edge state, which always keep their low-temperature features.

\subsection{Time-resolved charge current}

\subsubsection{Low and intermediate temperatures, $\alpha_\mathcal{E}\lesssim1$}

We find 
%
that, for a given period,
the time-resolved charge current, see, \eg, Ref.~\cite{Moskalets_book}, as function of the dimensionless time \mbox{$\tt=t/\sigma$} 
takes the form
\begin{equation}\label{eq_I_low_as}
	\frac{I_\text{cap/loc}(\tt)}{-e/\sigma}
	=
	\frac{D}{\pi}
	\left\{
	\frac{1}{\rule{0pt}{11pt}\tt^2+1}
	-
	\frac{1}{\left(\tt-\frac{\mathcal{T}}{2\sigma}\right)^{\!2}+1}
	\right\},
\end{equation}
with the electron-emission followed by the hole-emission after a half period. The specific shape stems from our choice for the emission times  \mbox{$t^{(\text{e})}=0$} (corresponding to \mbox{$\dcU=0$} for the driven capacitor). The shape of the current is Lorentzian with the width $\sigma$ for both emission schemes, see setups~A and~C in Fig.~\ref{fig_setup}. The time-resolved current~(\ref{eq_I_low_as}) is plotted as a solid red line in Fig.~\ref{fig_It}(a).
The leviton emission is described by the first (electron-emission) component, only; see the black dashed line in Fig.~\ref{fig_It}(a).

\subsubsection{High temperature, $\alpha_\mathcal{E}\gg1$}

\begin{figure}[t]
\includegraphics[scale=1]{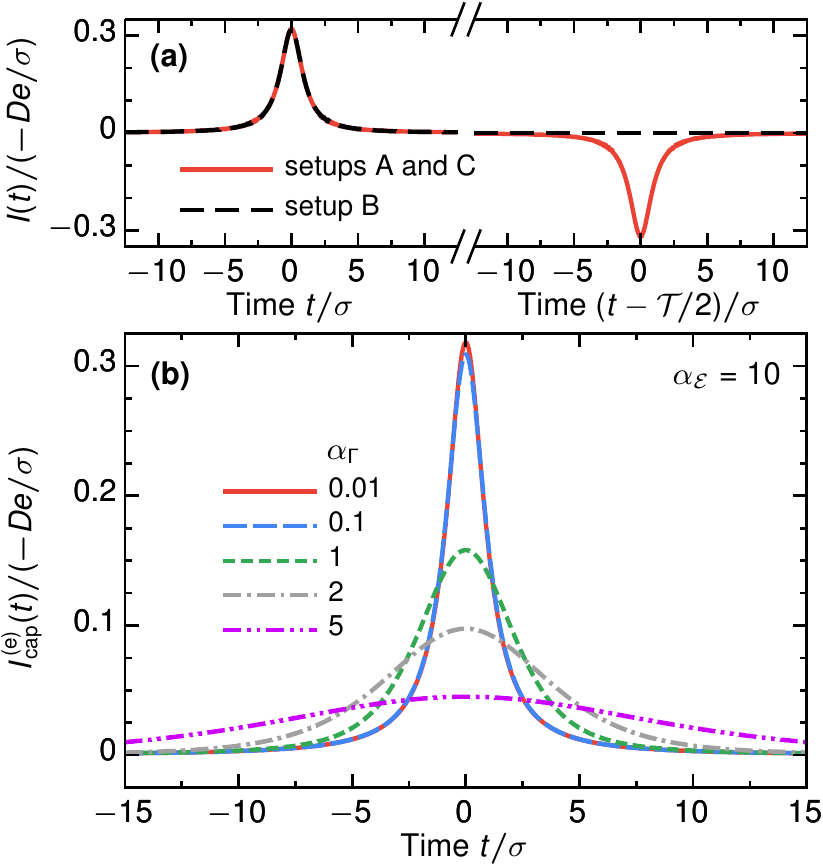}
\caption{\label{fig_It}
	Charge current as a function of time for \mbox{$\alpha_\mathcal{E}=10$} and \mbox{$\Omega\sigma\ll1$}. 
	Panel~(a) shows the charge currents emitted from the driven capacitor or the locally modulated edge state (setups~A and~C, respectively; see solid line) and from a Lorentzian-shaped bias voltage pulse (setup B; see dashed line) for relevant times of the whole driving period. Note the cut of the axis in the center. 
	Panel~(b) presents how the pulse emitted from the capacitor in the first half-period (\ie, the electron-emission contribution) changes with different values of $\alpha_\Gamma$, while the leviton pulse and the pulse emitted in setup~C remain unaltered.
}
\end{figure}

When the energy scale set by temperature exceeds the energy~$\mathcal{E}$ associated with the driving, the energy dependence of the scattering matrix of the mesoscopic capacitor can be resolved even in the limit of slow driving. In general, this yields important differences in the current properties. The time-resolved charge current emitted from the driven capacitor in this limit reads
\begin{align}\label{eq_I_high_as}
	\frac{I_\text{cap}(t)}{-e} 
	=\ &  
	D
	\!
	\sum_{j=\elec,\hole}
	\!
	\frac{\partial\Eresj(t)}{\partial t}\nonumber
\\
	&
	\times
	\int\!dE
	\bigg[-\frac{\partial f(E)}{\partial E}\bigg]
	\Lor_\Gamma \big(E-\Eresj(t)\big)
	\!
	.
\end{align}
One can see that the energy dependence of the scattering matrix, which is determined by the Lorentzian $\Lor_\Gamma$ with width $\Gamma$, is irrelevant as long as \mbox{$\alpha_\Gamma\ll1$}. Namely, in this limit, the result for the current is identical to the low-temperature result given in  Eq.~(\ref{eq_I_low_as}).

In contrast, the energy dependence of the scattering matrix becomes relevant, when the temperature reaches the level width, \mbox{$\alpha_\Gamma\geqslant1$}, demonstrating that it is exactly this factor $\alpha_\Gamma$ which determines the temperature dependence of the time-resolved current. For very large temperatures, $\alpha_\Gamma\gg1$, we find that the current is strongly suppressed and has a shape dictated by the derivative of the Fermi function,
\begin{eqnarray}
	I_\cap(t)
	=
	e D
	\sum_{j=\text{e,h}}
	\frac{\partial f\big(\Eresj(t)\big)}{\partial t}\ .
\end{eqnarray}
The role of the factor $\alpha_\Gamma$ can be seen best from the explicit result
\begin{equation}
	\frac{I_\text{cap}^{(\elec)}(\tt)}{-e/\sigma}
	=
	\frac{D}{\alpha_\Gamma}
	\cdot
	\frac{e^{\tt/\alpha_\Gamma}}{\left[1+e^{\tt/\alpha_\Gamma}\right]^2}\ .
\end{equation}
Here, for simplicity, we show the electron-emission contribution only. The hole-emission contribution  is given by an equivalent expression with an opposite sign.
The general result, obtained from carrying out the energy integration in Eq.~({\ref{eq_I_high_as}), is given in Appendix~\ref{app_general_It}. This general result is plotted in Fig.~\ref{fig_It}(b), where we show results for the time-resolved charge current in all temperature regimes. The suppression of the current for high temperatures, as well as the evolution of its characteristic shape from a Lorentzian to a derivative of a Fermi function become visible there.

\subsection{Time-resolved energy current}

\subsubsection{Low and intermediate temperatures, $\alpha_\mathcal{E}\lesssim1$}

The time-resolved energy current at low temperatures has very similar features as the charge current, Eq.~(\ref{eq_I_low_as}), where an energy $\mathcal{E}$ is transported per pulse instead of a charge $-e$ (or $e$ for holes),
\begin{equation}\label{eq_IE_low_as}
	\frac{I_{\cap/\loc}^\en(\tt)}{\en/\sigma}
	=
	\frac{2D}{\pi}
	\Bigg\{
	\frac{1}{\big[\tt^2+1\big]^{2}}
	+
	\frac{1}{\big[\big(\tt-\frac{\mathcal{T}}{2\sigma}\big)^{\!2}+1\big]^{2}}
	\Bigg\}
	.
\end{equation}
Note that both electron and hole pulses carry the same energy and that the pulse shape is sharper than the one of the charge current. Again, the leviton emission results in an energy current described by the first (electron-emission) contribution, only; see black dashed line in Fig.~\ref{fig_IEt}(a). The  result for setups A and C, Eq.~(\ref{eq_IE_low_as}), is plotted as a solid red line in Fig.~\ref{fig_IEt}(a).

%
\begin{figure}[t]
\includegraphics[scale=1]{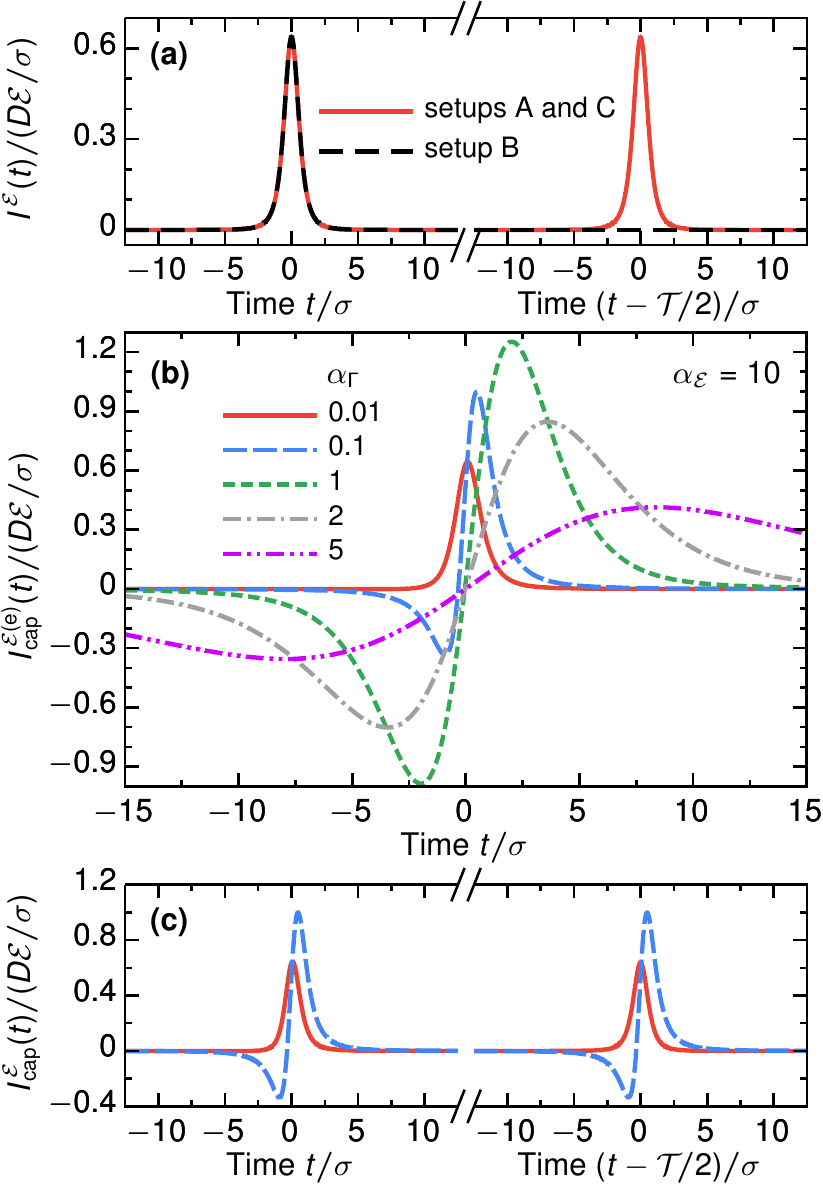}
\caption{\label{fig_IEt}
	Energy current as a function of time for \mbox{$\alpha_\mathcal{E}=10$} and \mbox{$\Omega\sigma\ll1$}. 
	Panels~(a) and~(b) are analogous to those shown in Fig.~\ref{fig_It} except that now the energy current is presented.
	Panel~(c) illustrates both electron- and hole-emission contributions to the energy current emitted  from the driven capacitor for two selected values of~$\alpG$ plotted in~(b).
}
\end{figure}

\subsubsection{High temperature, $\alpha_\mathcal{E}\gg1$}

The energy dependence of the scattering matrix of the driven mesoscopic capacitor, which gets important in the regime \mbox{$\alpha_\mathcal{E}\gg1$}, has a much stronger effect on the energy current than on the charge current. We find for the time-resolved energy current from the driven capacitor
\begin{align}\label{eq:IE_small_alps_fin}
	I_\cap^\en(t)
	=\ &
	D
	\!
	\sum_{j=\elec,\hole}
	\!
	\frac{\partial\Eresj(t)}{\partial t}
	\int\!d E
	\bigg[
	-
	\frac{\partial
	f(E)}{\partial E}
	\bigg]
\nonumber\\
	&
	\times	
	\bigg\{
	E
	\Lor_\Gamma\big(E-\Eresj(t)\big)	
\nonumber\\
	&
	+
	\pi\hbar	
	\frac{\partial\Eresj(t)}{\partial t}	
	\bigg[
	\big[\Lor_\Gamma\big(E-\Eresj(t)\big)\big]^2
\nonumber\\
	&
	+
	E
	\,
	\frac{\partial
	\big[\Lor_\Gamma\big(E-\Eresj(t)\big)\big]^2}{\partial E}
	\bigg]
	\bigg\}	
	.
\end{align}
Again, as long as \mbox{$\alpha_\Gamma\ll1$}, the same expression as Eq.~(\ref{eq_IE_low_as}) is obtained. However, when temperature is of the order of~$\Gamma$ or larger, the current gets significantly modified, see Appendix~\ref{app_general_IEt} for a full analytical expression. 
Equation~({\ref{eq:IE_small_alps_fin}) consists formally of two contributions: one that is antisymmetric and the other that is symmetric around the emission time. 
The antisymmetric part has a  shape similar to the time-resolved charge current, given in Eq.~(\ref{eq_I_high_as}). It can be understood as the energy carried by each particle, and hence, it changes its sign depending on whether the particle is injected above or below the reference electrochemical potential, \mbox{$\mu_\RR=0$}. Obviously, this term can only considerably contribute at large temperatures, where states both above and below the electrochemical potential are available for the single-particle emission. 
The second, symmetric part is the one that yields the low-temperature result for \mbox{$\alpha_\Gamma\ll1$}. The latter describes the energy injected into the conductor necessary for the excitation of a particle in the presence of the Fermi sea. Temperature also modifies this term, since the time at which this energy is injected into the conductor depends on the broadening of the level due to the coupling, $\Gamma$, and on temperature. 
See Appendix~\ref{app_IE_decompose} for a more detailed discussion of the two contributing terms. 
The characteristics of these two contributions become particularly evident in  the limit \mbox{$\alpha_\Gamma\gg1$}, where we find the insightful form of the current
\begin{align}\label{eq:IER_small_alpG}
\hspace*{-2pt}
	I_\cap^\en(t)
	=\ &
	D
	\sum_{j=\elec,\hole}
	\Eresj(t)
\nonumber\\
	&
	\times
	\bigg\{	
	\!
	-
	\frac{\partial
	f\big(\Eresj(t)\big)}{\partial t}
	+
	\frac{\hbar}{2\Gamma}
	\frac{\partial^2
	f\big(\Eresj(t)\big)}{\partial t^2}
	\bigg\}
	.
\end{align}
Which of these two terms dominates depends both on~$\alpG$ and on~$\alpen$. This is a fundamental difference with respect to the time-resolved charge current, which---at high temperatures---is governed exclusively by the coefficient~$\alpG$. It can be most clearly seen from the explicit evaluation of Eq.~(\ref{eq:IER_small_alpG}), leading to
\begin{multline}\label{eq:high-T_IEh}
	\frac{I_\cap^{\en(\elec)}(\tt)}{\en/\sigma}
	=
	D
	\frac{(\tt/\alpG)e^{\tt/\alpG}}{
	\big[1+e^{\tt/\alpG}\big]^2
	}
\\	
\times
	\frac{\Gamma}{\en}
	\bigg\{
	1
	+
	\frac{1}{\alpen}
	\tanh\bigg(\frac{\tt}{2\alpG}\bigg)
	\!
	\bigg\}
	,
\end{multline}
where for simplicity, we show only the electron-emission contribution.  
In Fig.~\ref{fig_IEt}(b), we display the time-resolved energy current for different values of~$\alpG$, starting from the general expression given in Appendix~\ref{app_general_IEt}. The more the temperature increases with respect to the level width~$\Gamma$, the more the properties of the time-resolved energy current change. The peak at the emission time, found for low temperatures, evolves into an anti-symmetric curve, which vanishes in the vicinity of the emission time. The energy transferred during the emission of electrons and holes is then almost fully governed by the temperature broadening of the Fermi sea: particles injected at negative/positive energies lead to negative/positive energy currents.\footnote{For the injection of a hole (absorption of an electron) in the second half of the cycle, an equivalent feature is found, see Fig.~\ref{fig_IEt}(c): particles absorbed at positive/negative energies lead to negative/positive energy currents.} The maximum and minimum of this close-to
antisymmetric curve is shifted to \mbox{$t\approx\pm\sigma\alpG$}. It means that with increasing temperature, particles can be emitted further and further away from the time at which the level crosses the Fermi energy of the reservoirs.
Note, however, that a small symmetric contribution always persists, guaranteeing that---also at high temperatures---the average energy carried per pulse is positive and given~by~$D\en$.

\section{Spectral currents}\label{sec_spec}

The energetic distribution of emitted  particles is  accessible \via the spectral current. It can be measured by inserting scatterers with  a specifically designed energy-dependent transmission into a setup~\cite{Altimiras2009Oct,Fletcher2013Nov,Battista2012}. The spectral current is given by
\begin{equation}\label{eq_spec_general}
	i(E)
	=
	D\sum_{n=-\infty}^\infty\left|S(E,E_n)\right|^2\left[f_\text{L}(E_n)-f_\text{R}(E)\right]\ .
\end{equation}
Here, for the bias-voltage driven system and the locally modulated edge state (setups B and C) the absolute valued squared of the scattering matrix has to be replaced by the energy-independent probabilities $|c_{\text{L},n}|^2$ from Eq.~(\ref{eq_cLn_resonant}) and $|c_{\text{g},n}|^2$ from Eq.~(\ref{eq_cgn_resonant}), respectively. 
However, also for the mesoscopic capacitor, we get from Eq.~(\ref{eq_S0_resonant_enrep}) for \mbox{$n\neq0$}
\begin{equation}\label{eq_Sn}
	\big|S_\cap(E,E_n)\big|^2
	=
	(2\Omega\sigma)^2
	e^{-2|n|\Omega\sigma} =: \left|S_n\right|^2
	,
\end{equation}
which is energy independent and equal to $\left| c_{\text{g},n}\right|^2$ and to $\left| c_{\text{L},n}\right|^2$ for \mbox{$n\geqslant0$}. 
It means that (in the regime considered here) the spectral currents of the three sources are almost identical at all temperatures, except for the fact that in the leviton case, only electrons are emitted. Technically, in Eq.~(\ref{eq_spec_general}), this difference arises from the vanishing contributions to $\left|c_{\text{L},n}\right|^2$ for negative $n$ and the constant part of the driving \mbox{$\dcVb=\hbar\Omega/(-e)$} entering the Fermi function $f_\text{L}(E_n)$, where it leads to an effective shift of the energy index \mbox{$f_\text{L}(E_n)\rightarrow f(E_{n-1})$}. 
This behavior  is in strong contrast to the time-resolved currents, which, as discussed in the previous section, were shown to have very different temperature-dependent characteristics. 

A general evaluation of the sum in Eq.~(\ref{eq_spec_general}), see Appendix~\ref{app_general_iE}, shows that the energy scale of the temperature only enters the spectral current \via the factor~$\alpen$.
Figure~\ref{fig_spectral} illustrates this evolution of the spectral current with changing~$\alpen$. The analytical results for the limiting cases for small and large temperatures elucidate the shape of the shown curves. In the limit of vanishingly small temperatures, \mbox{$\alpen\ll1$}, the spectral current as a function of the dimensionless energy \mbox{$\tE=E/\en$} is given~by
\begin{subequations}
\begin{align}
	i_\lev(\tE) 
	=\ &
	2\Omega\sigma\,D\, 
	e^{-|\tE|}
	\big[\theta(\tE)\big]
	,
\\
	i_{\cap/\loc}(\tE) 
	=\  &
	2\Omega\sigma\,D\,
	e^{-|\tE|}
	\big[\theta(\tE)-\theta(-\tE)\big]
	.
\end{align}
\end{subequations}
At low temperatures the shape of the spectral current is, thus, an exponential, see, \eg, Ref.~\cite{Moskalets_book}, with a width in energy given by $\en$. In the opposite limit of high temperatures, \mbox{$\alpen\gg1$}, we find
\begin{subequations} 
\begin{align}
	i_\lev(\tE) 
	=\ &  
	2\Omega\sigma\,D
	\bigg[
	\!-\frac{\partial}{\partial\tE}+2\frac{\partial^2}{\partial\tE^2}
	\bigg]
	\frac{1}{1+e^{\tE/\alpen}}
	,
\\
	i_{\cap/\loc}(\tE) 
	=\ &  
	2\Omega\sigma\,D
	\bigg[
	2\frac{\partial^2}{\partial\tE^2}
	\bigg]
	\frac{1}{1+e^{\tE/\alpen}}
	,
\end{align}
\end{subequations}
namely, a rounding of the signal with the shape dictated by derivatives of the Fermi function.

\begin{figure}[t]
\includegraphics[scale=1]{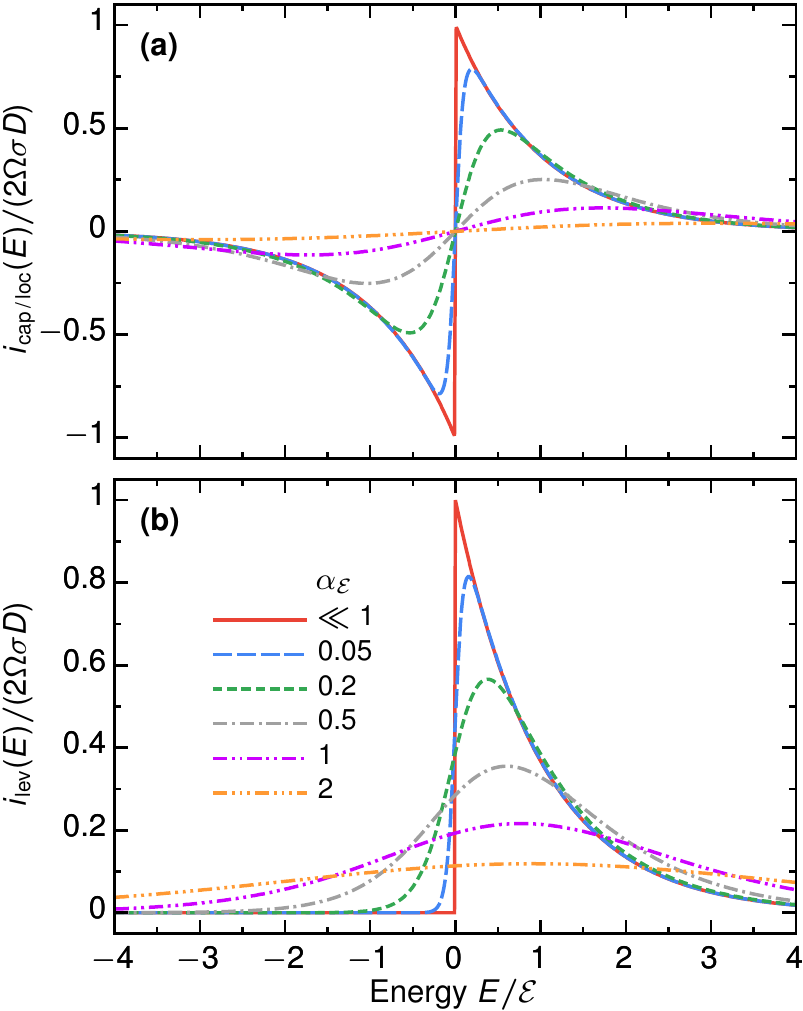}
\caption{\label{fig_spectral}
	Spectral current emitted  (a) from a time-de\-pen\-dent\-ly driven capacitor (setup A) or a locally modulated edge state (setup C), and (b)  from a Lorentzian-shaped bias voltage (setup B) as function of the dimensionless energy \mbox{$\tE=E/\en$} for different values of $\alpen$. 
}
\end{figure}

The conclusion from these results is that while the time-resolved current emitted from the bias driving and the local modulation of an edge state (setups B and C) are \emph{not} affected by temperature, the time-integrated, energy-resolved currents (\ie, the spectral currents) are actually strongly affected.
Nevertheless, the average energy transported by a single-particle pulse emitted from any of the three sources is not altered, 
$
	\int dt I^\en(t)
	=
	\en^2\mathcal{T}
	\int\!d\tE \tE i(\tE)
	=
	D\en
$, 
independently of temperature.

\section{Zero-frequency noise}\label{sec_noise}

In this section, we study zero-frequency noise of charge and energy currents, as well as the mixed charge-energy current noise.
In general, the study of correlation functions of this type yields additional information about a setup. It has been heavily exploited in the field of quantum optics with electrons based on single-electron sources. For example, the charge-current noise is a measure of the precision of the source~\cite{Reydellet03,Rychkov05,Vanevic2007Aug,Vanevic08,Albert2010Jul,Mahe2010Nov,Parmentier2012Apr,Moskalets2013Jul,Dubois13Theo,Vanevic2016Jan}, it has been used to identify correlations in Hanbury Brown-Twiss and Hong-Ou-Mandel setups or interferometer setups, where also the energy distributions of injected signals could be determined. 
Also the energy- (or heat-) current noise~\cite{Battista14,Moskalets2014May}, or even the mixed noise between charge and energy currents~\cite{Battista14a}, has recently been suggested as a tool to analyze setups with single-electron sources.  
Here, we investigate how these different types of noise are influenced by temperature effects and demonstrate that---except for the mixed noise---the results do not depend on the type of a source from which particles are injected.

\subsection{Charge-current noise}
\label{sec_PII}

The charge-current noise of the mesoscopic capacitor is found to be
\begin{align}\label{eq:PII_cap}
\hspace*{-4pt}
	\noise^{II}_\cap
	 =\ &  
	\frac{2e^2}{h}D^2
	\kB T	
\nonumber\\
	&\!+ 
	\frac{e^2}{h}
	D (1-D) 
	\!
	\sum_{n=-\infty}^\infty
	\!
	|S_n|^2
	n\hbar\Omega
	\coth\!\bigg(
	\!
	\frac{n\hbar\Omega}{2\kB T}
	\!
	\bigg)
	.
	\!
\end{align}
Note that only the energy-independent absolute value of the scattering matrix enters here. The same equation holds for the locally modulated edge state, where the absolute value squared of the scattering matrix is identical to the one of the driven mesoscopic capacitor. Employing an equivalent argument as in Sec.~\ref{sec_spec}, we find the same result for the charge-current noise of the leviton emission with the only difference that only positive $n$ contribute to the sum in Eq.~(\ref{eq:PII_cap}). 

\begin{figure}[t]
\includegraphics[scale=1]{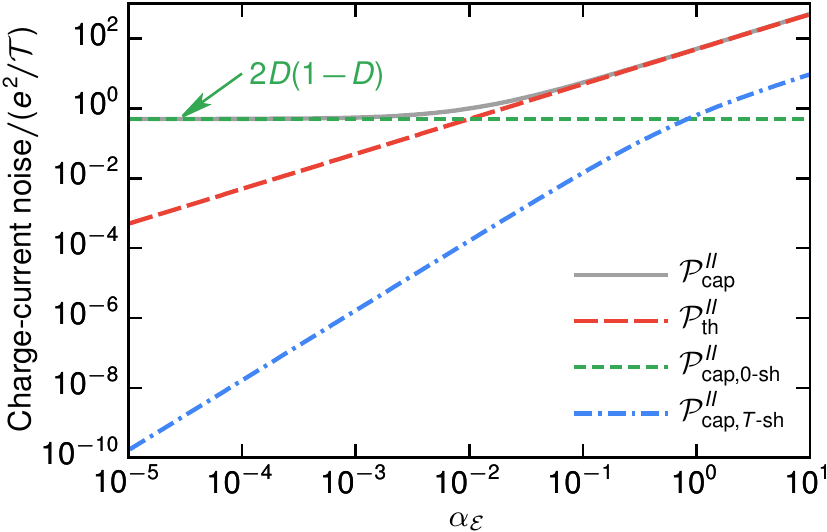}
\caption{\label{fig_charge_noise}
	Charge-current noise of the driven mesoscopic capacitor (setup A), which is identical to the one of setup~C.
	Importantly, for setup~B only the thermal contribution~$\noise_\text{th}^{II}$ is the same, whereas the driving-dependent contributions, $\noise_\text{0-sh}^{II}$ and~$\noise_\text{$T$-sh}^{II}$, differ by an overall factor 2.
	Three separate contributions, following Eq.~(\ref{eq:PII_split}), are shown additionally. We choose $D=0.5$ and $\Omega\sigma=0.01$.
}
\end{figure}

In the zero-temperature limit, this yields the well-known result for the shot noise~\cite{Reydellet03,Ol'khovskaya2008Oct}, 
\begin{subequations}\label{eq:PII}
\begin{equation}\label{eq:PII_0}
	\noise^{II}(\kB T\rightarrow0) 
	\equiv 
	\noise^{II}_\text{0-sh} 
	=	
	\frac{e^2}{\perT} 
	D(1-D)N
	,
\end{equation}
with  $N$ denoting the number of injected particles. This number equals \mbox{$N=1$} for the leviton-injection, whereas  \mbox{$N=2$} for the driven mesoscopic capacitor and the locally modulated edge state. For temperatures different from zero (namely, larger than the smallest energy-scale of the problem, \mbox{$\kB T>\hbar\Omega$}), we find 
\begin{equation}\label{eq:PII_split}
	\noise^{II}
	=
	\noise^{II}_\text{th}+\noise^{II}_\text{0-sh}+\noise^{II}_\text{$T$-sh}
	.
\end{equation}
The three contributions to Eq.~(\ref{eq:PII_split}) are the well-known thermal noise 
\begin{equation}\label{eq:PII_th}
	\noise^{II}_\text{th} 
	=
	\frac{2e^2}{h}D
	\kB T
	,
\end{equation}
which is independent of the driving, the zero-temperature shot noise, $\noise^{II}_\text{0-sh}$ as given in Eq.~(\ref{eq:PII_0}), and a third contribution  arising from shot noise at finite temperatures
\begin{equation}\label{eq:PII_Tsh}
	\noise^{II}_\text{$T$-sh} 
	=
	\frac{2e^2}{\perT} 
	D(1-D)	
	N	
	\Big[
	\alpen^2
	\psi^{(1)}(\alpen)
	-
	1
	\Big]
	,
\end{equation}
\end{subequations}
with $\psi^{(n)}(z)$ being the polygamma function of order~$n$.

In Fig.~\ref{fig_charge_noise}, we show the full charge-current noise~$\noise_\text{cap}^{II}$ for the mesoscopic capacitor (\ie, with \mbox{$N=2$}, solid gray line), as well as its separate contributions as decomposed in Eqs.~(\ref{eq:PII}). 
Except for the thermal noise, which is driving independent, $\noise_\text{cap}^{II}$ depends on temperature uniquely \via the parameter $\alpen$,\footnote{As expected, the broadening of the capacitor level is not a relevant energy-scale for the shot noise in the single-particle emission regime considered in this paper, since the (energy-dependent) emission times do not enter the noise expression.} which we choose as the plotting parameter here. 
As long as \mbox{$2\kB T<\hbar\Omega$} (or equivalently \mbox{$\alpen<\Omega\sigma$}), the thermal noise
plays no important role with respect to the zero-temperature shot noise, which is fully induced by the driving. The shot noise at finite temperatures exceeds the zero-temperature contribution only for \mbox{$\alpen>1$}. However, in the resonant limit, \mbox{$\Omega\sigma\ll 1$}, it takes place  solely in a regime, where the thermal noise is already the dominant noise contribution.

\subsection{Energy-current noise}

The energy-current noise of a time-dependently driven system, in general, consists of a transport contribution, which in the zero-temperature limit yields the shot noise, and an interference contribution~\cite{Dubois13Theo,Battista14}. The latter---in contrast to the interference part of the charge-current noise---depends on the driving and is nonzero even at zero temperatures. 
The general expression for the energy-current noise of the capacitor written in terms of scattering matrices, equivalent to Eq.~(\ref{eq:PII_cap}), is lengthy, and hence, included only in Appendix~\ref{app_deltat_contrib}. However, there, we also show that, in the single-particle emission regime considered here, both contributions are independent of the (energy-dependent) emission times of the scattering matrix. For this reason, also the energy-current noise of the three sources discussed in this paper differ only due to the number of injected particles. 
In the zero-temperature limit, one obtains
\begin{subequations}\label{eq:PEE}
\begin{align}
	\noise^{\en\en}(\kB T\rightarrow0)
	\equiv\ & 
	\noise^{\en\en}_\text{0-int}+\noise^{\en\en}_\text{0-sh}
\\\label{eq:PEE_lowT}
	=\ &
	\frac{\en^2}{\perT} 
	D^2N
	+
	\frac{\en^2}{\perT}
	2D(1-D)N
	,
\end{align}
where the first term is the zero-temperature interference contribution and the second one corresponds to the zero-temperature shot noise stemming from the transport part. 
%
\begin{figure}[b]
\includegraphics[scale=1]{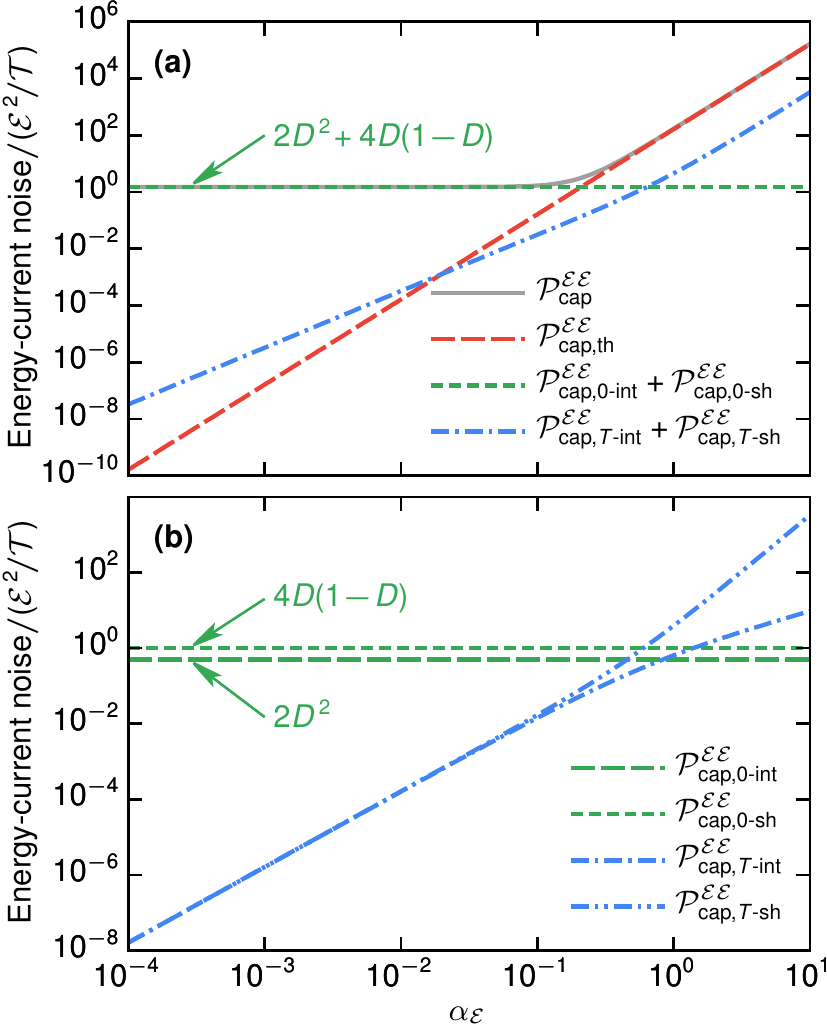}
\caption{\label{fig_energy_noise}
	Energy-current noise of the driven mesoscopic capacitor (setup A), which is identical to the one of setup~C.
	Importantly, for setup~B only the thermal contribution~$\noise_\text{th}^{\en\en}$ is the same, whereas all remaining, driving-dependent contributions differ by an overall factor 2.	
	Separate contributions, following Eqs.~(\ref{eq:PEE}), are shown additionally. Panel~(a) presents the pure thermal noise and the zero-temperature as well as the finite-temperature contribution of the combined transport and interference part. Panel~(b) illustrates the decomposition of the transport (shot) and interference part separately. We choose \mbox{$D=0.5$} and \mbox{$\Omega\sigma=0.01$}.
}
\end{figure}
%
When temperature gets larger (\mbox{$\kB T>\hbar\Omega$}), we find
\begin{equation}\label{eq:PEE_split}
	\noise^{\en\en} 
	=  
	\noise^{\en\en}_\text{th}
	+
	\noise^{\en\en}_\text{0-int}+\noise^{\en\en}_\text{0-sh}
	+
	\noise^{\en\en}_\text{$T$-int}+\noise^{\en\en}_\text{$T$-sh}.
\end{equation}
Thermal noise (namely, the driving-independent part of the interference contribution) is given by
\begin{equation}\label{eq:PEE_th}
	\noise^{\en\en}_\text{th} 
	=
	\frac{2\pi^2}{3h}D(k_\mathrm{B}T)^3
	. 
\end{equation}
In addition, both the interference and the transport part get driving-dependent thermal contributions
\begin{align}\label{eq:PEE_T}
	\noise^{\en\en}_\text{$T$-int}
	=\ &
	\frac{2\en^2}{\perT}
	D^2N
	\Big[
	\alpen^2
	\psi^{(1)}(\alpen)
	-
	1
	\Big]
	,		
\\
	\noise^{\en\en}_\text{$T$-sh}
	=\ &
	\frac{2\en^2}{3\perT}
	D(1-D)N
	\Big\{
	\alpen^4
	\psi^{(3)}(\alpen)
	-
	6	
\nonumber\\				
	&+	
	\pi^2
	\alpen^2
	\Big[
	\alpen^2
	\psi^{(1)}(\alpen)
	-
	1/2
	\Big]	
	\Big\}
	.
\end{align}
\end{subequations}
In Fig.~\ref{fig_energy_noise}, we show the full energy-current noise, as well as its separate contributions. The main contribution to the total energy-current noise [plotted in panel~(a) as the solid gray line] is given by the sum of  zero-temperature transport and interference parts [plotted in panel~(a) as the finely dashed green line] up to temperatures fulfilling the condition \mbox{$2(\kB T)^3\approx\en^2/\perT$}, or equivalently \mbox{$\alpen\approx\sqrt[3]{\Omega\sigma}$}. It means that the parameter $\alpen$, at which thermal noise starts to play a role is much larger than for the charge-current noise. The driving-dependent thermal contributions for the energy current become relevant at \mbox{$\alpen\approx1$}, but stay well below the pure thermal noise. This behavior is analogous to that found for the charge-current noise. 
Finally, all the transport and interference contributions are plotted separately in Fig.~\ref{fig_energy_noise}(b). One can observe there that the transport part is much more affected by temperature than the interference part. We note that a separate readout of transport and interference contributions is enabled by their different $D$-dependence.

\subsection{Mixed noise}

Last but not least, we present results for the mixed zero-frequency correlator of charge and energy currents. 
The general expression for the mixed noise from the driven capacitor and the locally modulated edge state is given by
\begin{equation}\label{eq:PIE_lev}
\hspace*{-5pt}
	\noise^{I\en}_\cap
	= 
	\frac{-e}{2h}
	D (1-D) 
	\!
	\sum_{n=-\infty}^\infty
	\!
	|S_n|^2
	\left(n\hbar\Omega\right)^2
	\coth\!\bigg(
	\!
	\frac{n\hbar\Omega}{2\kB T}
	\!
	\bigg)
	\!
	.
	\!
\end{equation}
Due to the pure ac driving of setups~A and~C, going along with the emission of an equal number of electrons and holes, the mixed noise vanishes identically, \mbox{$\noise_{\cap/\loc}^{I\en}\equiv 0$}.
Technically, this arises from the fact that the first two factors in the sum in Eq.~(\ref{eq:PIE_lev}) are even, see also Eq.~(\ref{eq_Sn}), while the third one is odd. Instead, for the leviton emission~\cite{Battista14a}, as realized in setup~B, one finds a finite contribution due to the asymmetry induced by the pure electron emission (or equivalently, by the nonvanishing zero-component of the bias).
In the zero-temperature limit, the shot noise contribution to the mixed noise reads
\begin{subequations}\label{eq:PIE}
\begin{equation}
	\noise^{I\en}(\kB T\rightarrow0)
	\equiv
	\noise^{I\en}_\text{0-sh}
	=
	\frac{-e\en}{\perT}
	D(1-D)\delta N
	,
\end{equation}
with the difference between injected electrons and holes: \mbox{$\delta N=1$} (setup~B) and \mbox{$\delta N=0$} (setup~A and C).
For finite temperatures (\mbox{$\kB T>\hbar\Omega$}), there are only two additional contributions to the mixed noise, 
\begin{equation}\label{eq:PIE_split}
	\noise^{I\en}
	=
 	\noise^{I\en}_\text{0-sh}
 	+
 	\noise^{I\en}_\text{$T$-sh}+\noise^{I\en}_\text{$T$-int}
 	.
\end{equation}
We see that there is no \textit{pure} thermal contribution. The full interference contribution is both temperature- and driving-dependent,
\begin{equation}
	\noise^{I\en}_\text{$T$-int} 
	=
	\frac{e^2}{h}D^2\delta N\dcVb k_\text{B}T
	=
	\frac{-e\en}{\perT}
	D^2\delta N\alpen
	,
\end{equation}
with \mbox{$\dcVb=\hbar\Omega/(-e)$}. The thermal contribution to the shot noise (transport part) is given by
\begin{equation}
	\noise^{I\en}_\text{$T$-sh} 
	=
	\frac{-e\en}{\perT}
	D(1-D)\delta N
	\left[-\alpen^3\psi^{(2)}(\alpen)-1\right]
	.
\end{equation}
\end{subequations}
Figure~\ref{fig_mixed_noise} presents the full mixed noise and its separate contributions. As for the other noise expressions, the driving-dependent thermal transport and interference parts of $\noise^{I\en}_\text{lev}$ contribute with equal weight starting from~\mbox{$\alpen\approx 1$}.

\begin{figure}[t]
\includegraphics[scale=1]{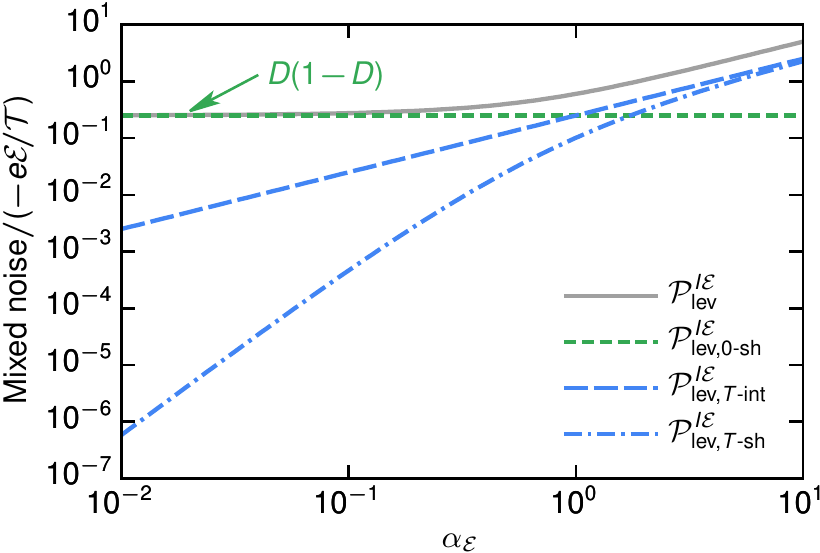}
\caption{\label{fig_mixed_noise}
	Mixed charge-energy current noise for the Lorentz\-ian-shaped bias voltage driving (setup~B) leading to emission of levitons.
	We choose \mbox{$D=0.5$} and \mbox{$\Omega\sigma=0.01$}.
}
\end{figure}

\section{Conclusions and Outlook}

We have analyzed the characteristics of time-resolved charge and energy, as well as spectral currents, together with the zero-frequency correlators of charge and energy currents of single-electron sources, which emit noiseless single-particle pulses as minimal excitations of the Fermi sea. 
The main focus is on how these observables vary for different types of single-electron source. In particular, the difference in temperature dependence of the observables is analyzed.
The relevant types of single-electron sources, which we consider in this paper are as follows: (setup A) a driven mesoscopic capacitor  emitting single particles due to a pure ac gate driving of a confined region with a discrete energy-spectrum; (setup B) leviton emission  taking place due to a time-dependent bias voltage (with a non-vanishing dc component~$\dcVb$) in the absence of any confinement in the conductor; (setup~C) an emission scheme not relying on the confinement but allowing for electron and hole emission due to a local modulation of an edge state.

The inherent properties of these sources lead to the following observable effects in the analyzed quantities.
First of all, due to the energy dependence of the scattering matrix of the driven capacitor, resulting from its discrete spectrum, the time-resolved charge and energy currents for such a source are strongly temperature-dependent. In particular, the ratio between the driving-independent level broadening~$\Gamma$, Eq.~(\ref{eq_gamma}), and temperature plays an important role here. This behavior is in contrast to that for the leviton emission and the locally modulated edge state, where temperature has no impact on time-resolved currents.

Furthermore, when looking at spectral currents and zero-frequency noises, the energy dependence of the scattering matrix plays no role. Instead, it is the energy dependence of the Fermi functions, which entirely determines temperature-dependent effects. 
Major differences in the observed quantities for different sources  result then from the number of emitted particles and from the spectral weight of particles at positive and negative energies (differing for electrons and holes). 
Interestingly, the mixed charge-energy current noise has a special role: it requires a non-vanishing dc~component of the bias $\dcVb$, and thus, it vanishes identically in setups~A and~C, where electrons are emitted from a pure ac gate-driving. 
The temperature dependence of the spectral current and noises is fully governed by the ratio between driving dependent quantities, namely the frequency $\Omega$ or the energy emitted per pulse $\en$, as compared to temperature.

These fundamental properties of the major observables used for analysis in quantum optics with electrons, which can be ascribed to variations of a small set of relevant parameters ($\Omega\sigma$, $\alpG$, and $\alpen$), are expected to be important for future studies of setups containing any of the three single-particle sources analyzed in this paper. Specifically, it will help to distinguish complex features\mbox{---deriving,} for instance, from correlations between different particles---from the fundamental source properties.

\acknowledgments

Funding from the Knut and Alice Wallenberg Foundation through the Academy Fellow program (J.S., N.D. and M.M.), from the EU ITN
PhD4Energy, Grant No. 608153 (S.K.), and from the Swedish VR is gratefully acknowledged.

\appendix

\section{Currents and noise---definitions}\label{app_general_exp}

In this appendix, we define the charge and energy current operators, their expectation values, and their correlations. These expressions can also be found in textbooks and reviews, such as, Refs.~\cite{Blanter00,Moskalets_book}. 
For the charge and energy current operators in contact $\alpha$  we have
\begin{equation}\label{eq_I_operator}
	\opI_\alpha(t)  
	=  
	-\frac{e}{h}
	\int_{-\infty}^\infty\!\!dE dE^\prime
	\,
	\skew{2}{\hat}{i}_\alpha(E,E^\prime) 
	\,
	e^{i(E-E^\prime)t/\hbar}
	,
\end{equation}
with \mbox{$e>0$}, as well as 
\begin{equation}\label{eq_IE_operator}
	\opI^\en_\alpha(t)  
	=  
	\frac{1}{h}\int_{-\infty}^\infty\!\!dE dE^\prime 
	\frac{E+E^\prime}{2}
	\,
	\hat{i}_\alpha(E,E^\prime)
	\,
	e^{i(E-E')t/\hbar}
	.
\end{equation}
The operator 
\mbox{$
	\opi_\alpha(E,E^\prime)
	=
	\hat{b}^\dagger_\alpha(E)\hat{b}_\alpha(E^\prime)
	-
	\hat{a}^\dagger_\alpha(E) \hat{a}_\alpha(E^\prime)
$} 
contains creation and annihilation operators for incoming [\mbox{$\hat{b}^\dagger_\alpha(E)\hat{b}_\alpha(E^\prime)$}] and outgoing [\mbox{$\hat{a}^\dagger_\alpha(E) \hat{a}_\alpha(E^\prime)$}] fluxes in contact~$\alpha$. For simplicity, we restrict the present analysis to the case of a single transport channel, which can be extended to the general multi-channel case, \eg, as done in Ref.~\cite{Battista14}.
The incoming and outgoing scattering states are related to each other through a scattering matrix
\begin{equation}
	\hat{b}_{\alpha}(E_n)
	=
	\sum_{\beta=\LL,\RR}
	\,
	\sum_{m=-\infty}^{\infty} 
	\!
	S_{\alpha \beta}(E_n,E_m)
	\hat{a}_{\beta}(E_m)
	,
\end{equation}
with \mbox{$E_m\equiv E+m\hbar\Omega$}, and $m\hbar\Omega$ (for \mbox{$m\in \mathbb{Z}$}) standing for the amount of absorbed or emitted energy quanta in a scattering process. See the main text for specific expressions.

In order to obtain the time-resolved charge and energy currents, $I_\alpha(t)$ and $I^\en_\alpha(t)$, the expectation values of Eqs.~(\ref{eq_I_operator}) and~(\ref{eq_IE_operator}), respectively,  have to be calculated
\begin{equation}\label{eq_Itime_general}
	I_\alpha(t)
	=
	\langle\opI_\alpha(t)\rangle
	\quad
	\text{and}
	\quad
	I^\en_\alpha(t)
	=
	\langle\opI^\en_\alpha(t)\rangle	
	,
\end{equation}
where \mbox{$\langle\ldots\rangle$} denotes the quantum statistical average.
Moreover, the time-averaged charge and energy currents, $\avI_\alpha$ and $\avI^\en_\alpha$, can be derived \via integration of Eq.~(\ref{eq_Itime_general}) within one period $\perT$, that is,
\mbox{$
	\avI_\alpha
	=
	\int_0^\perT\!\!dt
	\langle\opI_\alpha(t)\rangle/\perT
$} 
and 
\mbox{$
	\avI^\en_\alpha
	=
	\int_0^\perT\!\!dt
	\langle\opI^\en_\alpha(t)\rangle/\perT
$}.

The fluctuations of charge and energy currents are accessible through the (generalized) zero-frequency noise, 
\begin{equation}\label{eq_noise_general}
	\noise^{XY}_{\alpha\beta}
	=
	\int^{\mathcal{T}}_{0}\!\dfrac{dt}{\mathcal{T}} 
	\int^{\infty}_{-\infty}\!dt^\prime
	\,
	\langle \Delta X_{\alpha}(t) \Delta Y_{\beta}(t+t^\prime) \rangle 
	.
\end{equation}
The current fluctuation, \mbox{$\Delta \hat{X}(t)=\hat{X}(t)-\langle\hat{X}(t)\rangle $}, is kept general here, and will in the following be replaced by the fluctuations of the relevant current operators, $\opI$ or $\opI^\en$.

\section{General expressions for currents and noises}\label{app_general_results}

Here, we present general expressions for the currents and their correlators describing a situation where both time-dependent driving of the contacts, as well as a time-dependent scatterer, such as the driven mesoscopic capacitor or the locally modulated edge state, can be present in the same setup. For this reason, we incorporate both the Floquet scattering matrix of the central region, $S_{\alpha\beta}(E_n,E_m)$, as well as coefficients accounting for the driving of the \textit{bias} voltage,~$c_{\alpha,m}$ with $\alpha=\text{L,R}$. 
Employing Eqs.~(\ref{eq_I_operator})-(\ref{eq_IE_operator}), we find the general expression for the time-resolved charge and energy currents to be:
\begin{equation}\label{eq_It_general_1}
	I_\alpha^{(\en)}(t)
	=
	\sum_{n=-\infty}^\infty
	e^{-in\Omega t}
	\,
	I_{\alpha,n}^{(\en)}
\end{equation}
with\footnote{Note that we make the following assumption for current (and noise) measurements in setups with time-dependent bias driving: the detection of the signal is always assumed to take place \textit{before} particles scattered at the central scattering region enter again the region of time-dependent bias driving.}
\begin{align}\label{eq_It_general_2}
\hspace*{-7pt}
\renewcommand{\arraystretch}{1.35}
	\begin{pmatrix}
	I_{\alpha,n}
	\\
	I_{\alpha,n}^\en
	\end{pmatrix}
	\!=&\
	\frac{1}{h}
	\sum_\beta
	\sum_{m,q=-\infty}^\infty
	\int\!dE
	\begin{pmatrix}
	-e
	\\
	E+\frac{n\hbar\Omega}{2}
	\end{pmatrix}
\nonumber\\
	&
	\times
	\!
	\Big\{
	S_{\alpha\beta}^\ast(E,E_m)
	S_{\alpha\beta}(E_n,E_{m+q})
	\,
	\auxF_{\beta,q}(E_m)
\nonumber\\
	&	
	\hspace*{5pt}
	-
	S_{\alpha\beta}^\ast(E,E_m)
	S_{\alpha\beta}(E_{n-q},E_m)
	\,
	\auxF_{\alpha,q}(E)
	\Big\}
	,
	\!
	\!
\end{align}
with the auxiliary function~$\auxF_{\alpha,q}(E)$ defined as 
\begin{equation}\label{eq:auxF}
	\auxF_{\alpha,q}(E)
	=
	\sum_{l=-\infty}^\infty
	\!
	c_{\alpha,l}^\ast
	c_{\alpha,l+q}^{}
	f_\alpha(E_{-l})
	.
\end{equation}
Note that~$\auxF_{\alpha,q}(E)$ has a clear physical interpretation only for \mbox{$q=0$},  representing then an effective non-equilibrium distribution function induced by the ac driving in contact~$\alpha$. This function,
\begin{equation}\label{eq_eff_distrib}
	\FDeff_\alpha(E)
	\equiv
	\auxF_{\alpha,0}(E)
	=	
	\sum_{l=-\infty}^\infty
	\!
	|c_{\alpha,l}|^2	
	f_\alpha(E_{-l})
	,
\end{equation} 
takes into account the fact that in the presence of the ac driving also states originating from energies, which differ from $E$ by an integer multiple of~$\hbar\Omega$, contribute to transport at energy $E$~\cite{Battista14}.

Furthermore, from Eqs.~(\ref{eq_It_general_1})-(\ref{eq_It_general_2}) one can obtain  the time-averaged currents, corresponding to the term for \mbox{$n=0$} in the series~(\ref{eq_It_general_1}),
\begin{equation}
	\avI_\alpha^{(\en)}
	=
	\int_0^\perT\!\frac{dt}{\perT}
	\,
	I_\alpha^{(\en)}(t)
	=
	I_{\alpha,0}^{(\en)}
	.
\end{equation}
An equivalent procedure for deriving the time-averaged currents involves calculation of energy integrals over a spectral current~$i_\alpha(E)$, that is,
\mbox{$
	\avI_\alpha 
	=
	-e\int\!dE
	\,
	i_\alpha(E)/h
$},
and 
\mbox{$
	\avI^\en_\alpha 
	=
	\int\!dE
	\,
	E i_\alpha(E)/h
$}, 
with
\begin{multline}\label{eq_Ispectral_general}
	i_\alpha(E) 
	= \sum_\beta
	\sum_{m,q=-\infty}^\infty 
	S_{\alpha\beta}^\ast(E,E_m)
	S_{\alpha\beta}^{}(E,E_{m+q})
\\
	\times
	\big[
	\auxF_{\beta,q}(E_m)
	-
	\delta_{q0}\,
	\auxF_{\alpha,q}(E)
	\big]
	.
\end{multline}
While the previous equations for the current operators, Eqs.~(\ref{eq_I_operator})-(\ref{eq_IE_operator}), were kept on a very general level, here we have explicitly evaluated expectation values of the occupations of incoming current states, leading to the occurrence of the Fermi function 
$
	f_\alpha(E)
	=
	\big\{
	1+\exp[(E-\mu_\alpha)/(\kB T)]
	\big\}^{-1}
$.

Finally,  we also express the noise~(\ref{eq_noise_general}) in terms of Floquet scattering matrices and Fermi functions: 
\begin{widetext}
\begin{align}\label{eq:Pab_def}
	\mathcal{P}_{\alpha\beta}^{XY}
	=&\ 
	\frac{1}{h}
	\int\!\!dE
	\sum_{q,l,k=-\infty}^\infty
	\!
	x
	y_{q+l}
	\,
	\Big\{
	\delta_{\alpha\beta}
	\,
	\delta_{q0}
	\,
	\delta_{k0}
	\,
	\auxF_{\alpha,l}(E)
	\big[
	\delta_{l0} 
	- 
	\auxF_{\alpha,-l}(E_l)	
	\big]
	-
	S_{\alpha\beta}^\ast(E,E_q)
	S_{\alpha\beta}^{}(E,E_{q+k})
\nonumber\\[-2pt]
	&\, \times
	\auxF_{\beta,l}(E_q)
	\big[
	\delta_{kl} 
	- 
	\auxF_{\beta,k-l}(E_{q+l})	
	\big]
-
	S_{\beta\alpha}^\ast(E_{q+l},E_l)
	S_{\beta\alpha}^{}(E_{q+l},E_k)
	\,
	\auxF_{\alpha,k}(E)
	\big[
	\delta_{l0} 
	- 
	\auxF_{\alpha,-l}(E_l)	
	\big]
\nonumber\\[-2pt]
	&\,
	+
	\sum_{\gamma\lambda}
	\sum_{n,m=-\infty}^\infty
	\!
	S_{\alpha\gamma}^\ast(E,E_n)
	S_{\alpha\lambda}^{}(E,E_{m+k})
	S_{\beta\lambda}^\ast(E_{q+l},E_m)
	S_{\beta\gamma}^{}(E_{q+l},E_{n+l})
	\,
	\auxF_{\gamma,l}(E_n)
	\big[
	\delta_{k0} 
	- 
	\auxF_{\lambda,k}(E_m)	
	\big]	
	\Big\}
	.
\end{align}
%
%
In the equation above we use a compact notation for the charge-current~($\noise_{\alpha\beta}^{II}$), energy-current~($\noise_{\alpha\beta}^{\en\en}$) and mixed noise ($\noise_{\alpha\beta}^{I\en}$ and $\noise_{\alpha\beta}^{\en I}$), which can be obtained by doing the replacements \mbox{$x\rightarrow\{-e,E\}$} and \mbox{$y_{q+l}\rightarrow\{-e,E_{q+l}\}$} for \mbox{$X, Y\rightarrow\{I,\en\}$}. Notice the notation used for distinguishing  different types of noise where, for the sake of brevity, the superscript `$\en$' refers to the energy current~$I^\en$.

In order to relate these general expressions to the three setups discussed in the main text, the Floquet scattering matrices $S_{\alpha\beta}(E_n,E_m)$ have to be replaced by elements of the following matrices:
\begin{equation}\label{eq_scat_mat_cap}
\renewcommand{\arraystretch}{1.35}
	\bm{S}_\sA(E_n,E_m) 
	=
	\begin{pmatrix}
	\sqrt{1-D}\ S_\cap(E_n,E_m) & \delta_{nm}\sqrt{D}\\
	\sqrt{D}\ S_\cap(E_n,E_m) & -\delta_{nm}\sqrt{1-D}
	\end{pmatrix}
	\!
\end{equation}
for the setup with the driven mesocopic capacitor, 
\begin{equation}\label{eq_scat_mat_lev}
\renewcommand{\arraystretch}{1.35}
	\bm{S}_\sB(E_n,E_m)
	=
	\delta_{nm}
	\begin{pmatrix}
	\sqrt{1-D} & \sqrt{D}\\
	\sqrt{D} & -\sqrt{1-D}
	\end{pmatrix}
\end{equation}
for the time-dependent bias voltage driving and %
\begin{align}\label{eq_scat_mat_loc}
	\bm{S}_\sC(E_n,E_m) 	
\renewcommand{\arraystretch}{1.5}
	=
	\begin{pmatrix}
	\sqrt{1-D}\ 	
	e^{iE_m\taug/\hbar}
	c_{\gate,n-m}	 & \delta_{nm}\sqrt{D}\\
	\sqrt{D}\ 
	e^{iE_m\taug/\hbar}
	c_{\gate,n-m}    & -\delta_{nm}\sqrt{1-D}
	\end{pmatrix}
	\!
\end{align}
for the locally modulated edge state.
Furthermore, in the cases A and C, namely, in the absence of a bias driving, only the amplitudes $c_{\LL,0}$ and \mbox{$c_{\RR,0}$} contribute to the observables discussed in this section.

\section{Analytic results for time-resolved currents}

\subsection{Charge current}
\label{app_general_It}

The high-temperature (\mbox{$\alpen\gg1$}) expression for the charge current~$I_\cap(t)$ for arbitrary~$\alpG$ is obtained by performing explicitly the energy integration in Eq.~(\ref{eq_I_high_as}), which yields
\begin{align}
	I_\cap(t)
	=\ &
	\frac{eD}{2\pi}
	\sum_{j=\elec,\hole}
	\big[
	\partial_t
	\Eresj(t)
	\big]
	\Big\{
	\pi
	\partial_E
	\funA(E)\big|_{E=\Eresj(t)}
	+ 
	\frac{\beta}{2\pi}
	\text{Re}\Psi_-^{(1)}
	\!
	\Big(
	\frac{\beta}{2\pi}
	\big[\Eresk(t)+i\Gamma\big]
	\Big)
	\Big\}
	.
\end{align}
For the sake of notational brevity, in the equation above and henceforth we use \mbox{$\beta\equiv1/(\kB T)$}, and also employ the short notation $\partial_x\equiv\partial/\partial x$ for derivatives.
The auxiliary functions
are defined as
\begin{equation}\label{eq_Psi_def}
	\Psi_\pm^{(n)}(z)
	=
	\psi^{(n)}(1/2+iz)
	\pm
	\psi^{(n)}(1/2-iz)
	\hspace{0.5cm}\text{and}\hspace{0.5cm}
	\funA(E)
	=
	2
	\text{Re}
	f(E+i\Gamma)
	,
\end{equation}
with $\psi^{(n)}(z)$ denoting the polygamma function of order~$n$.
Below we provide the explicit expression for the electron-emission contribution to the charge current, 
\begin{align}
	\frac{I_\cap^{(\elec)}(\tt)}{-e/\sigma}
	=\ &
	\frac{D}{2\pi\alpG}
	\Bigg\{
	\pi
	\frac{
	1+\cosh(\tt/\alpG)\cos(1/\alpG)
	}{\big[\cosh(\tt/\alpG)+\cos(1/\alpG)\big]^2}
	-
	\frac{1}{2\pi}
	\text{Re}\Psi_-^{(1)}
	\!
	\Big(
	\frac{1}{2\pi\alpG}
	\big[\tt+i\big]
	\Big)
	\Bigg\}
	,
\end{align}
where the time~$t$ is now scaled to the temporal width~$\sigma$, that is, $\tt=t/\sigma$, and we have used Eq.~(\ref{eq_tE}) for the emission energy~$\Eresel(t)$, which implies our choice of $\tcape(E=0)=0$. This equation is used to plot results shown in Fig.~\ref{fig_It}.

\subsection{Energy current}
\label{app_general_IEt}

Analogously as in the previous section, one can also derive the high-temperature (\mbox{$\alpen\gg1$}) expression for the energy current~$I^\en_\cap(t)$ for arbitrary~$\alpG$ by calculating in Eq.~(\ref{eq:IE_small_alps_fin}) the integrals with respect to energy, 
\begin{equation}
	I_\cap^\en(t)
	=
	\sum_{j=\elec,\hole}
	\Big\{	
	I_{\cap,\sym}^{\en(j)}(t)
	+
	I_{\cap,\asym}^{\en(j)}(t)
	\Big\}
	,
\end{equation}
where
%
%
\begin{align}\label{eq_IE_s}
	I_{\cap,\sym}^{\en(j)}(t)
	=
	-
	\frac{D}{2\pi}	
	\big[\partial_t\Eresj(t)]
	\Big\{	
	&
	\Eresj(t)	
	\Big[
	\pi
	\,
	\partial_E
	\funA(E)\big|_{E=\Eresj(t)}
	+
	\frac{\beta}{2\pi}
	\text{Re}\Psi_-^{(1)}
	\!
	\Big(
	\frac{\beta}{2\pi}
	\big[\Eresj(t)+i\Gamma\big]
	\Big)
	\Big]
\nonumber\\	
	+\ &
	\Gamma
	\Big[
	\pi
	\,
	\partial_E
	\funB(E)\big|_{E=\Eresj(t)}	
	-
	\frac{\beta}{2\pi}
	\text{Im}\Psi_-^{(1)}
	\!
	\Big(
	\frac{\beta}{2\pi}
	\big[\Eresj(t)+i\Gamma\big]
	\Big)
	\Big]
	\Big\}
	,
\end{align}
and
\begin{align}\label{eq_IE_as}
	I_{\cap,\asym}^{\en(j)}(t)
	=
	\frac{D\hbar}{4\pi}	
	\big[\partial_t\Eresj(t)]^2	
	\Big\{	
	&
	\frac{1}{\Gamma}
	\Eresj(t)
	\Big[
	\pi
	\,
	\partial_E^2
	\funA(E)\big|_{E=\Eresj(t)}
	-
	\frac{\beta^2}{(2\pi)^2}	
	\text{Im}\Psi_+^{(2)}
	\!
	\Big(
	\frac{\beta}{2\pi}
	\big[\Eresj(t)+i\Gamma\big]
	\Big)	
	\Big]
\nonumber\\
	+\ &
	\Gamma
	\Big[
	\pi
	\,
	\partial_E^3
	\funA(E)\big|_{E=\Eresj(t)}	
	-
	\frac{\beta^3}{(2\pi)^3}		
	\text{Re}\Psi_-^{(3)}
	\!
	\Big(
	\frac{\beta}{2\pi}
	\big[\Eresj(t)+i\Gamma\big]
	\Big)		
	\Big]
\nonumber\\
	-\ &
	\Eresj(t)
	\Big[
	\pi
	\,
	\partial_E^3
	\funB(E)\big|_{E=\Eresj(t)}
	+
	\frac{\beta^3}{(2\pi)^3}		
	\text{Im}\Psi_-^{(3)}
	\!
	\Big(
	\frac{\beta}{2\pi}
	\big[\Eresj(t)+i\Gamma\big]
	\Big)				
	\Big]	
	\Big\}
	.
\end{align}
Recall that the auxiliary functions~$\Psi_\eta^{(n)}(z)$ and~$\funA(E)$ are given by Eqs.~(\ref{eq_Psi_def}), while
\begin{equation}
	\funB(E)
	=
	-2
	\text{Im}
	f(E+i\Gamma)
	.
\end{equation}
Furthermore, one can show that, for instance, the electron-emission contributions to the energy  current take the explicit form (with \mbox{$\tt\equiv t/\sigma$}):
\begin{align}\label{eq_IEt_s_elec}
	\frac{I_{\cap,\sym}^{\en(\elec)}(\tt)}{\en/\sigma}
	=
	\frac{D}{2\pi}	
	\cdot
	\frac{\Gamma}{\en}
	\bigg\{	
	&
	\frac{\tt}{\alpG}
	\bigg[
	\pi
	\,
	\frac{
	1+\cosh\big(\tt/\alpG\big)\cos(1/\alpG)
	}{
	\big[\cosh\big(\tt/\alpG\big)+\cos(1/\alpG)]^2}	
	-
	\frac{1}{2\pi}
	\text{Re}\Psi_-^{(1)}
	\!
	\Big(
	\frac{1}{2\pi\alpG}
	\big[\tt+i\,\big]
	\Big)
	\bigg]
\nonumber\\	
	+\ &
	\frac{1}{\alpG}
	\bigg[
	\pi
	\,
	\frac{
	\sinh\big(\tt/\alpG\big)\sin(1/\alpG)
	}{
	\big[\cosh\big(\tt/\alpG\big)+\cos(1/\alpG)]^2}		
	+
	\frac{1}{2\pi}
	\text{Im}\Psi_-^{(1)}
	\!
	\Big(
	\frac{1}{2\pi\alpG}
	\big[\tt+i\,\big]
	\Big)
	\bigg]
	\bigg\}
	,
\end{align}
and
\begin{align}\label{eq_IEt_as_elec}
	\frac{I_{\cap,\asym}^{\en(\elec)}(\tt)}{\en/\sigma}
	=
	\frac{D}{2\pi}	
	\bigg\{	
	&
	\frac{\tt}{\alpG^2}
	\bigg[
	\pi
	\,
		\sinh\big(\tt/\alpG \big)
		\frac{
		2-\cos^2(1/\alpG)+\cosh\big(\tt/\alpG\big)\cos(1/\alpG)
		}{
		\big[\cosh\big(\tt/\alpG\big)+\cos(1/\alpG)]^3}	
	-
	\frac{1}{(2\pi)^2}	
	\text{Im}\Psi_+^{(2)}
	\!
	\Big(
	\frac{1}{2\pi\alpG}
	\big[\tt+i\,\big]
	\Big)	
	\bigg]
\nonumber\\
	+\ &
	\frac{1}{\alpG^3}
	\bigg[
	\pi
	\,
	\mathcal{C}_{\mathcal{A}}(\tt,\alpG)	
	-
	\frac{1}{(2\pi)^3}		
	\text{Re}\Psi_-^{(3)}
	\!
	\Big(
	\frac{1}{2\pi\alpG}
	\big[\tt+i\,\big]
	\Big)		
	\bigg]
\nonumber\\	
	-\ &
	\frac{\tt}{\alpG^3}		
	\bigg[
	\pi
	\,
	\mathcal{C}_{\mathcal{B}}(\tt,\alpG)	
	+
	\frac{1}{(2\pi)^3}		
	\text{Im}\Psi_-^{(3)}
	\!
	\Big(
	\frac{1}{2\pi\alpG}
	\big[\tt+i\,\big]
	\Big)				
	\bigg]	
	\bigg\}
	,
\end{align}
where the auxiliary functions~$\mathcal{C}_{\mathcal{A}}(\tt,\alpG)$ and $\mathcal{C}_{\mathcal{B}}(\tt,\alpG)$ are given by
\begin{align}
	\mathcal{C}_{\mathcal{A}}(\tt,\alpG)	
	=\ &
	\frac{
	1
	}{
	\big[\cosh\big(\tt/\alpG \big)+\cos(1/\alpG)]^4}
	\bigg\{
	3
	-
	\cosh\big(2\tt/\alpG\big)
	+
	2
	\sinh^2\big(\tt/\alpG\big)	
	\cos(2/\alpG)
\nonumber\\
&
	-
	\frac{1}{2}
	\cosh\big(\tt/\alpG\big)\cos(1/\alpG)
	\big[
	-6
	+
	\cos(2/\alpG)
	+
	\cosh\big(2\tt/\alpG\big)
	\big]
	\bigg\}	
	,
\end{align}
\begin{align}
	\mathcal{C}_{\mathcal{B}}(\tt,\alpG)	
	=\ &
	-
	\sinh\big(\tt/\alpG \big)\sin(1/\alpG)
	\frac{
	\cos(2/\alpG)
	+
	\cosh\big(2\tt/\alpG\big)
	-
	8\cosh\big(\tt/\alpG \big)\cos(1/\alpG)
	-
	10
	}{
	2\big[\cosh\big(\tt/\alpG\big)+\cos(1/\alpG)]^4}
	.
\end{align}
These equations are used to plot results shown in Fig.~\ref{fig_IEt} as well as Fig.~\ref{fig_IE_split}.
%

\section{Temperature dependence of the time-resolved energy current}
\label{app_IE_decompose}

In this brief appendix, we separately present plots of the antisymmetric and symmetric contribution of the time-resolved energy current of the capacitor, $I_\text{cap,s}^\en(t)$ and $I_\text{cap,as}^\en(t)$, in the high-temperature limit, as given in Eq.~(\ref{eq:IE_small_alps_fin}) and explicitly evaluated in Eqs.~(\ref{eq_IE_s}) and~(\ref{eq_IE_as}).  
For simplicity, we only show the electronic part, \cf Eqs.~(\ref{eq_IEt_s_elec}) and~(\ref{eq_IEt_as_elec}); the hole emission yields equivalent results shifted by one half of the period.
Panels (a) and (b) in the left column of Fig.~\ref{fig_IE_split} show the behavior for fixed $\alpen=10$ for different values of $\alpG$. We see that with increasing $\alpG$, an antisymmetric contribution arises, which is fully absent in the zero-temperature limit. For $\alpG\simeq1$ the absolute value of the maximum/minimum values of this contribution stop increasing. Subsequently, merely a broadening, going along with a shift of the time at which the maximum/minimum occurs, \mbox{$t_\text{max}\approx\pm\sigma\alpG$}, can be observed. 

\begin{figure*}[t]
\includegraphics[scale=1]{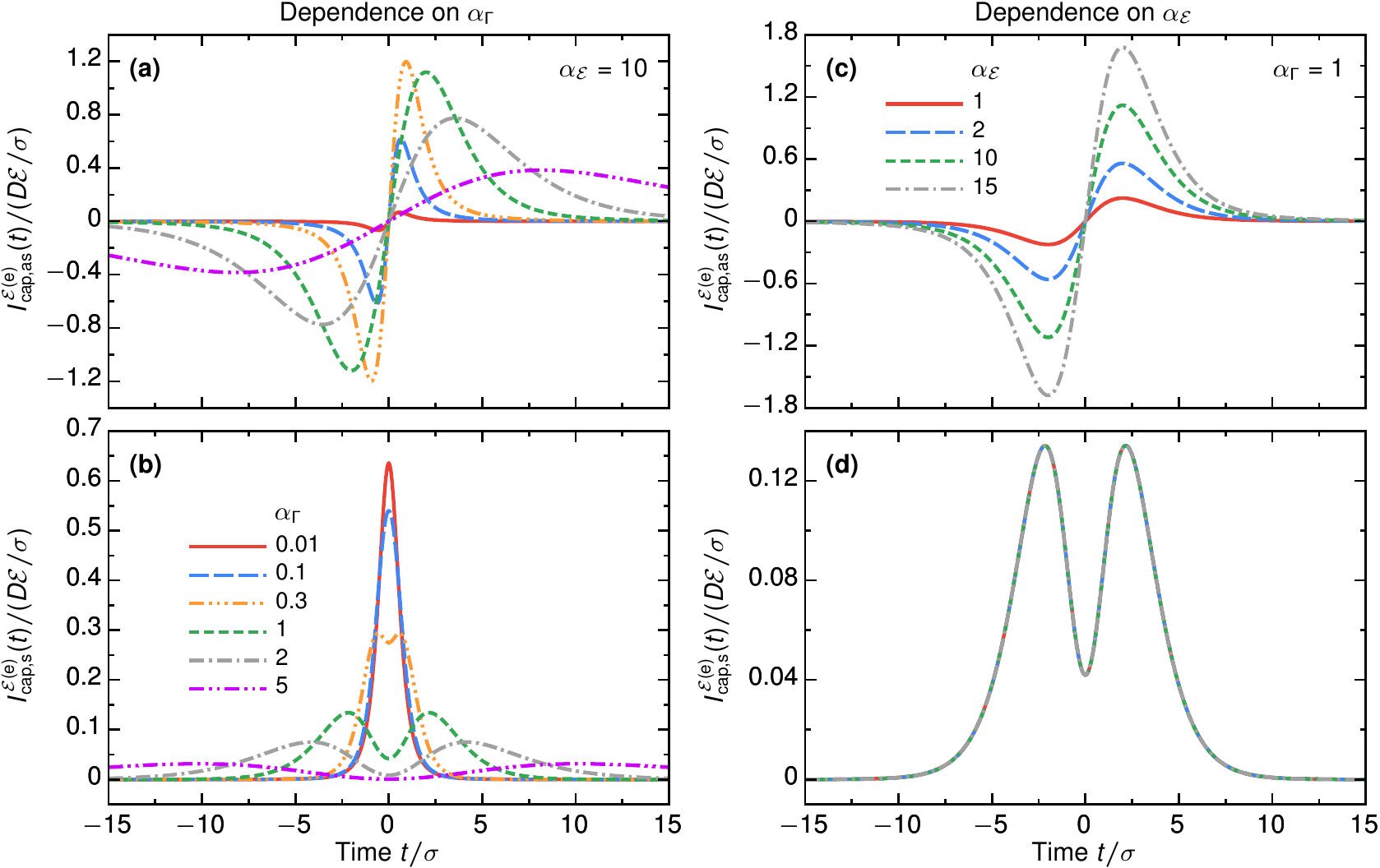}
\caption{\label{fig_IE_split}
	Electron-contribution to the time-resolved energy current emitted from the driven mesoscopic capacitor (setup~A) in the high-temperature limit, \mbox{$\alpen>1$}. Here, we present the antisymmetric $I_\text{cap,as}^\en(t)$ and symmetric $I_\text{cap,s}^\en(t)$ contributions to Eq.~(\ref{eq:IE_small_alps_fin}) separately.
}
\end{figure*}

The symmetric component, $I_\text{cap,s}^\en(t)$, shown in panel (b) has a very different behavior as function of $\alpG$. The peak at \mbox{$t=0$}, which is observed at zero temperature, splits into two peaks, again with maxima occurring at approximately \mbox{$t_\text{max}\approx\pm\alpG\sigma$}, and with the temperature broadening. The integral over the whole curve remains constant with changing $\alpG$.

Interestingly, the influence of $\alpen$, shown in panels (c) and (d) in the right column of Fig.~\ref{fig_IE_split} barely results in an overall increase of the weight of the antisymmetric part of the function $I^\en_\text{cap}(t)$. The symmetric contribution is completely independent of the factor $\alpen$.

This behavior reflects that explicitly presented in Eq.~(\ref{eq:high-T_IEh}) for \mbox{$\alpG\gg1$}. Note, however, that the plots shown in this appendix are valid for arbitrary $\alpG$.

\section{Analytic results for spectral currents}
\label{app_general_iE}
\subsection{Time-dependently driven mesoscopic capacitor (setup A) and locally modulated edge state (setup~C)}

In the case of the time-dependently driven mesoscopic capacitor and the locally modulated edge state, one employs Eq.~(\ref{eq_S0_resonant_enrep}) for $\big|S(E,E_n)\big|^2$ and the equivalent expression from Eq.~\eqref{eq_Sloc} [in combination with Eq.~(\ref{eq_cgn_resonant})], respectively, so that Eq.~(\ref{eq_spec_general}) reads
\begin{equation}\label{eq_iEcap_sum_aux}
	i_{\cap/\loc}(E)
	=
	(2\Omega\sigma)^2
	D
	\sum_{n=0}^\infty 
	e^{-2n\Omega\sigma}
	\big[
	f(E_n)
	+
	f(E_{-n})
	-
	2f(E)
	\big]
	.	
\end{equation}
Next, in the limit \mbox{$\alpen\gg1$}  considered here,\footnote{Note that, strictly speaking, the condition \mbox{$\kB T\gg\hbar\Omega$}, which is weaker than $\alpen\gg1$, is sufficient for the transformation from the sum into an integral.} 
one can convert the sum in Eq.~(\ref{eq_iEcap_sum_aux}) into an integral over $
	\omega
	=
	n\hbar\Omega/(\kB T)
$,
\begin{equation}\label{eq_iEcap_integral_aux}
	i_{\cap/\loc}(E)
	=
	2\Omega\sigma
	\, 
	D
	\,
	\alpen
	\int_0^\infty\!d\omega
	\,
	e^{-\omega\alpen}
	\big[
	f(E+\omega\kB T)
	+
	f(E-\omega\kB T)
	-
	2f(E)
	\big]
	.
\end{equation}
From there we obtain with the help of the definition of the (Gauss) hypergeometric function~$\hgf(a,b,c;z)$ \cite{Abramowitz_book},
\begin{align}\label{eq_iEcap_general}
\hspace*{-3pt}
	i_{\cap/\loc}(\tE)
	=\ &
	2\Omega\sigma
	D
	\bigg\{
	\tanh\!\bigg(
	\!
	\frac{\tE}{2\alpen}
	\!
	\bigg)
	+
	\tildehgf\big(\alpen;\text{e}^{-\tE\!/\alpen}\big)
	-
	\tildehgf\big(\alpen;\text{e}^{\tE\!/\alpen}\big)
	\bigg\}
	,
\end{align}
where the dimensionless energy \mbox{$\tE=E/\en$} has been introduced, and
\begin{equation}\label{eq_tildeghf_def}
	\tildehgf\big(\alpen;z\big)
	=
	\frac{\alpen}{1+\alpen}
	z
	\hgf\big(1,1+\alpen,2+\alpen;-z\big)
	.
\end{equation}
Results corresponding to Eq.~(\ref{eq_iEcap_general}) are plotted in Fig.~\ref{fig_spectral}(a).

\subsection{Lorentzian bias driving (setup B)}

In order to evaluate the sum in Eq.~(\ref{eq_spec_general}) for the spectral current in the case of the Lorentzian bias driving, let us first replace $\big|S(E,E_n)\big|^2$ by $|c_{\text{L},n}|^2$ given by Eq.~(\ref{eq_cLn_resonant}) and use that \mbox{$\dcVb=\hbar\Omega/(-e)$},
\begin{align}\label{eq_iElev_sum_aux}
	i_\lev(E)
	=\ &
	(2\Omega\sigma)^2
	D
	\sum_{n=0}^\infty 
	e^{-2n\Omega\sigma}
	\big[
	f(E_n)
	-
	f(E)
	\big]
	.	
\end{align}
Then,  as discussed in the previous subsection, one can write
\begin{align}\label{eq:ilev_integral_aux}
\hspace*{-3pt}
	i_\lev(E)
	=\ &
	2\Omega\sigma
	\,
	D
	\,
	\alpen
	\int_0^\infty\!d\omega
	\,
	e^{-\omega\alpen}
	\big[
	f(E+\omega\kB T)
	-
	f(E)
	\big]
	.
\end{align}
With the help of the hypergeometric function~$\hgf(a,b,c;z)$, the equation above can be shown to have the following form
\begin{equation}\label{eq_iElev_general}
	i_\lev(\tE)
	=
	2\sigma\Omega
	\,
	D
	\bigg[
	\frac{1}{1+\text{e}^{-\tE/\alpen}}
	-
	\tildehgf\big(\alpen;\text{e}^{\tE/\alpen}\big)
	\bigg]
	.
\end{equation}
Equation~(\ref{eq_iElev_general}) has been used to plot results shown in Fig.~\ref{fig_spectral}(b).

\section{Derivation of the energy-current noise }\label{app_deltat_contrib}

In this appendix, we derive the general expressions for the noises in setup A, based on a mesoscopic capacitor. Using Eqs.~(\ref{eq:Pab_def}) and (\ref{eq_scat_mat_cap}) in the absence of a bias driving \mbox{$\auxF_{\alpha,q}(E)=\delta_{q,0}f_\alpha(E)$}, we find
%
\begin{align}\label{eq_noise_cap_general}
	\noise_\cap^{XY}
	=
	\frac{1}{h}
	\sum_{n=-\infty}^\infty
	\!
	\bigg\{
	&
	D(1-D)
	\!
	\int\!\!dE
	\,
	xy	
	\,
	\big|S_\cap(E,E_{-n})\big|^2
	A_{\LL\RR}^n(E)
+
	D^2
	\bigg[
	\delta_{n0}
	\int\!\!dE
	\,
	x y
	\,
	F_{\RR\RR}^{0,0}(E)
\nonumber\\[-5pt]
	+\ &
	\sum_{k,l=-\infty}^\infty
	\int\!\!dE
	\,
	x y_{n}
	\,
	S_\cap^*(E,E_{-l})
	S_\cap(E_n,E_{-l})
	S_\cap^*(E_n,E_{-k})
	S_\cap(E,E_{-k})
	\,
	F_{\LL\LL}^{l,k}(E)
	\bigg]	
	\bigg\}
	,
\end{align}
%
where we have introduced two auxiliary functions~$F_{\alpha\beta}^{n,m}(E)$ and~$A_{\alpha\beta}^q(E)$ defined as
\begin{equation}\label{eq_aux_fun_F}
	F_{\alpha\beta}^{n,m}(E)
	\equiv
	f_\alpha(E_{-n})\big[1-f_\beta(E_{-m})\big]
	\hspace{0.5cm}\text{and}\hspace{0.5cm}
	A_{\alpha\beta}^q(E)
	\equiv
	F_{\alpha\beta}^{q,0}(E)
	+	
	F_{\beta\alpha}^{0,q}(E)
.
\end{equation}
%
%
%
First of all, it can be seen that formally Eq.~(\ref{eq_noise_cap_general}) consists of two types of terms---the first \mbox{$\propto D(1-D)$} and the second \mbox{$\propto D^2$}---which have a clear physical interpretation~\cite{Dubois13Theo,Battista14}. Specifically, the term \mbox{$\propto D(1-D)$} describes correlations generated by the exchange of particles  between two different reservoirs, and hence, it is often referred to as the \emph{transport part} of the noise. Conversely, the term \mbox{$\propto D^2$}, the so-called \emph{interference part} of the noise,  represents correlations due to the exchange of particles between states with different energies in the same reservoir.

Unlike for the charge-current noise, see Sec.~\ref{sec_PII}, where the dependence of the scattering matrix~$S_\cap(E_n,E_m)$ on energy effectively did not play a role, for the energy-current noise this dependence has to be carefully taken into account. 
In particular, from Eq.~(\ref{eq_noise_cap_general}) we derive 
\begin{align}\label{eq_PEE_cap}
\hspace*{-2pt}
	\noise^{\en\en}_\cap
	=\ &
	\frac{D (1-D)}{3h}
	\sum_{n=-\infty}^\infty
	\!
	|S_n|^2
	\coth\!\bigg(
	\!
	\frac{n\hbar\Omega}{2\kB T}
	\!
	\bigg)
	\big[
	\big(n\hbar\Omega\big)^{\!3}
	+
	n\hbar\Omega
	\big(\pi\kB T\big)^{\!2}
	\big]
\nonumber\\
	&\!
	+
	\frac{D^2}{h}
	\bigg\{
	\frac{2\pi^2}{3}
	(\kB T)^3
	+
	\sum_{n>0}^\infty
	\coth\!\bigg(
	\!
	\frac{n\hbar\Omega}{2\kB T}
	\!
	\bigg)	
	\int\!\!dE
	\,
	\big|\Lambda_n(E)\big|^2
	\big[
	f(E)-f(E_n)
	\big]
	\bigg\}
	,
\end{align}
where
\begin{equation}\label{eq_Lambda_n_def}
	\Lambda_n(E)
	=
	\sum_{k=-\infty}^\infty
	\!
	k\hbar\Omega
	\,
	S_\cap^\ast(E_k,E)
	S_\cap(E_k,E_n)	
	.
\end{equation}
Comparing the equation above to the analogous equation for the Lorentzian bias driving (\cf Eqs.~(13) and~(15) of Ref.~\cite{Battista14}), it can be noticed that apart from the lack of a bias voltage applied to contacts, one finds in Eq.~(\ref{eq_PEE_cap}) an additional term---originating from the interference part of Eq.~(\ref{eq_noise_cap_general})---which arises due to the energy dependence of the scattering matrix.
Replacing the scattering matrices in Eq.~(\ref{eq_Lambda_n_def}) by their explicit form~(\ref{eq_S0_resonant_enrep}) [recall that $|S_n|^2$ in Eq.~(\ref{eq_PEE_cap}) is given by Eq.~(\ref{eq_Sn})], and finding that
\begin{equation}
	\big|\Lambda_n(E)\big|^2
	=
	2(\hbar\Omega)^2
	\,
	e^{-2n\Omega\sigma}
	\big\{
	1
	-
	\cos\!\big[
	n\Omega\Delta t(E)
	\big]
	\big\}		
	,
\end{equation}
with
$
	\Delta t(E)
	\equiv
	\tel(E)-\tho(E)
$,
we eventually obtain
\begin{align}\label{eq_PEE_cap_2}
\hspace*{-7pt}
	\noise^{\en\en}_\cap
	=\ &
	\frac{2\pi^2}{3h}
	D
	(\kB T)^3
	+
	\frac{2D^2}{h}
	(\hbar\Omega)^2	
	\sum_{n>0}^\infty	
	e^{-2n\Omega\sigma}
	n\hbar\Omega
	\coth\!\bigg(
	\!
	\frac{n\hbar\Omega}{2\kB T}
	\!
	\bigg)
\nonumber\\
	&\!
	+
	\frac{D (1-D)}{3h}
	(2\Omega\sigma)^2
	\sideset{}{^\prime}\sum_{n=-\infty}^\infty
	\!
	e^{-2|n|\Omega\sigma}
	\coth\!\bigg(
	\!
	\frac{n\hbar\Omega}{2\kB T}
	\!
	\bigg)
	\big[
	\big(n\hbar\Omega\big)^{\!3}
	+
	n\hbar\Omega
	\big(\pi\kB T\big)^{\!2}
	\big]
	+
	\big(\noise^{\en\en}_\cap\big)^{\!\prime}
	.
\end{align}
Here, the additional term~$\big(\noise^{\en\en}_\cap\big)^{\!\prime}$ is given by
\begin{align}\label{eq_PEEprime_cap}
\hspace*{-8pt}
	\big(\noise^{\en\en}_\cap\big)^{\!\prime}
	\equiv\ &
	-
	\frac{2D^2}{h}
	(\hbar\Omega)^2
	\sum_{n>0}^\infty
	(-1)^n
	\,
	e^{-2n\Omega\sigma}
	\coth\!\bigg(
	\!
	\frac{n\hbar\Omega}{2\kB T}
	\!
	\bigg)	
	\auxI_n
	,
\end{align}
with 
\begin{equation}\label{eq_energy_integral}
	\auxI_n	
	=
	\int\!\!dE
	\,
	\cos\!\big[
	2n\Omega\tel(E)
	\big]	
	\big\{
	f(E)-f(E_n)
	\big\}	
	=
	\pi\kB T
	\,
	\sin\!\bigg(\!2n^2\dfrac{\hbar\Omega}{e\acU}\!\bigg)
	\sinh^{-1}\!\bigg(\!2\pi n\dfrac{\kB T}{e\acU}\bigg)
	.
\end{equation}
Here, we have used that
\mbox{$
	\cos[
	n\Omega\Delta t(E)
	]	
	=
	(-1)^n
	\cos[
	2n\Omega\tel(E)
	]		
$}, from Eq.~(\ref{eq_tcap_energy}).
Importantly, the term~$\big(\noise^{\en\en}_\cap\big)^{\!\prime}$ can be conveniently estimated for the parameter regime considered in this paper by invoking the conditions~(\ref{eq_res_cond}) and~(\ref{eq_temp_cond}), \ie, \mbox{$\Omega\sigma\ll1$} and \mbox{$\kB T/|e\acU|\ll1$}, and noting that the energy of the  Floquet quantum, $\hbar\Omega$, is in fact the smallest energy scale under the adiabatic-driving condition, meaning that \mbox{$\hbar\Omega/(\kB T)\ll1$}. 
In order to analyze $	\big(\noise^{\en\en}_\cap\big)^{\!\prime}$, let us insert Eq.~(\ref{eq_energy_integral}) into Eq.~(\ref{eq_PEEprime_cap}), which yields
\begin{equation}
\hspace*{-2pt}
	\frac{\big(\noise^{\en\en}_\cap\big)^{\!\prime}}{\en^2/\perT}	
	=
	-4D^2
	\,
	\sigma\Omega
	\,\alpen
	\sum_{n>0}^\infty
	(-1)^n
	\,
	a_n
	\quad
	\text{with}
	\quad
	a_n
	=
	\pi
	e^{-2n\Omega\sigma}	
	\coth\!\bigg(
	\!
	\frac{n\hbar\Omega}{2\kB T}
	\!
	\bigg)		
	\sin\!\bigg(\!2n^2\dfrac{\hbar\Omega}{\kB T}\cdot\dfrac{\kB T}{e\acU}\!\bigg)
	\sinh^{-1}\!\bigg(\!2\pi n\dfrac{\kB T}{e\acU}\bigg)
	.
	\!
\end{equation}
Then, the key task is reduced to the calculation of the series 
\mbox{$	
	\sum_{n>0}^\infty
	(-1)^n
	\,
	a_n
$}.
This series is convergent (as it is absolutely converging, which can be shown from the comparison test). Furthermore, in the parameter range mentioned above~$a_n$ and~$a_{n+1}$ differ very little for all~$n$. We can therefore approximate this series as follows
\begin{equation}
	\sum_{n>0}^\infty
	(-1)^n
	\,
	a_n
	\approx
	-
	\sum_{k=0}^\infty
	\frac{da_n}{dn}\bigg|_{n=2k+1}
	\approx
	-\frac{1}{2}
	\int_0^\infty\!dn
	\,
	\frac{da_n}{dn}
	=
	1
	,
\end{equation}
so that we obtain 
\begin{equation}
	\frac{\big(\noise^{\en\en}_\cap\big)^{\!\prime}}{\en^2/\perT}	
	=
	-4D^2
	\,
	\sigma\Omega
	\,\alpen
	.
\end{equation}
Consequently, we can conclude that under the conditions considered in this paper, $\big(\noise^{\en\en}_\cap\big)^{\!\prime}$ is negligibly small as compared to the remaining terms in Eq.~(\ref{eq_PEE_cap_2}), see also Eqs.~(\ref{eq:PEE}) for explicit expressions.

\end{widetext}


%

\end{document}